\documentclass[conference]{./IEEEtran}

\usepackage[available,functional,reproduced]{ndssbadges}

\usepackage{listings}

\usepackage{graphicx}
\usepackage{subfigure}
\graphicspath{{./figs/}{./figs/}}
\DeclareGraphicsExtensions{.eps, .pdf, .mps}

\usepackage{amsmath}
\usepackage{mathtools}
\usepackage{nccmath}
\usepackage{amssymb}
\usepackage{tabulary}
\usepackage{booktabs}
\usepackage{tabularx}
\usepackage{soul}

\usepackage{lipsum}

\usepackage{hyperref} \hypersetup{colorlinks=true, urlcolor=blue, linkcolor=blue, citecolor=red}
\usepackage{breakurl}
\usepackage{cleveref}
\usepackage{algorithm}
\usepackage{algpseudocode}
\algnewcommand{\LineComment}[1]{\Statex {\color{red} {/* \textit{#1} */}}}

\usepackage{xspace}
\usepackage{tikz}
\tikzset{%
	fake reference/.style={
		rounded corners,
		inner sep=6pt
	}}
\usepackage{easyReview}

\newcommand{\eg}{{\em e.g.}, }
\newcommand{\ie}{{\em i.e.}, }
\newcommand{\alg}{{\scshape Iris}\xspace}
\newcommand{\prv}{{\slshape $(\alpha,\delta)$-privacy}\xspace}
\newcommand{\prvt}{{\slshape $(\alpha,\delta)$-private}\xspace}
\newcommand{\dnodes}{\nu}
\newcommand{\dobjects}{\kappa}

\newcommand{\reqexample}{44}
\newcommand{\ttt}[1]{\texttt{#1}}

\usepackage{amsthm}
\newtheorem{definition}{Definition}

\usepackage{balance}

\newcommand\eatpunct[1]{}

\begin{document}

\title{Iris: Dynamic Privacy Preserving Search in Authenticated Chord Peer-to-Peer Networks}

\author{\IEEEauthorblockN{Angeliki Aktypi}
	\IEEEauthorblockA{University of Oxford\\
		angeliki.aktypi@cs.ox.ac.uk}
	\and
	\IEEEauthorblockN{Kasper Rasmussen}
	\IEEEauthorblockA{University of Oxford\\
		kasper.rasmussen@cs.ox.ac.uk}}


\maketitle

\begin{abstract}
In structured peer-to-peer networks, like Chord, users find data by
asking a number of intermediate nodes in the network. Each node
provides the identity of the closet known node to the address of the
data, until eventually the node responsible for the data is reached.
This structure means that the intermediate nodes learn the address of
the sought after data. Revealing this information to other nodes makes
Chord unsuitable for applications that require query privacy so in
this paper we present a scheme \alg to provide query privacy while
maintaining compatibility with the existing Chord protocol. This means
that anyone using it will be able to execute a privacy preserving
query but it does not require other nodes in the network to use it (or
even know about it).

In order to better capture the privacy achieved by the iterative
nature of the search we propose a new privacy notion, inspired by
$k$-anonymity. This new notion called \prv, allows us to formulate
privacy guarantees against adversaries that collude and take advantage
of the total amount of information leaked in all iterations of the
search.

We present a security analysis of the proposed algorithm based on the
privacy notion we introduce. We also develop a prototype of the
algorithm in Matlab and evaluate its performance. Our analysis proves
\alg to be \prvt while introducing a modest performance
overhead. Importantly the overhead is tunable and proportional to the
required level of privacy, so no privacy means no overhead.

\end{abstract}


\IEEEpeerreviewmaketitle

\section{Introduction}
\label{sec:intro}
Structured Peer-to-Peer (P2P) networks,
provide a scalable and robust lookup service
allowing a requester to identify the provider of sought-after information.
The Chord~\cite{stoica2001chord} lookup service is one
of the first structured P2P networks to be widely deployed, and has been used in systems such as
the Cooperative File System (CFS)~\cite{frank2001cfs},
UsenetDHT~\cite{sit2008usenetdht}, OverCite~\cite{stribling2006overcite} and ConChord~\cite{ajmani02conchord}.
It has also been proposed in the literature as a resource service discovery mechanism in grid computing~\cite{pirro2016chordgrid}
and WSN~\cite{labbai2016t2wsn} and an alternative to the traditional centralised design for
DNS~\cite{russ2002dns} and telephony~\cite{kundan2005telephony} systems.
More recently, CFS has been proposed
by the Tor project~\cite{syverson2004tor} as an efficient key-value lookup system with authenticated updates
to allow the retrieval of the introductory points for onion services.
Chord also underpins the NKN (New Kind of Network)~\cite{nkn2018whitepaper}
blockchain network infrastructure focused on decentralising network resources
used as the base for many decentralised applications (dapps)
including nMobile, \mbox{D-chat}, nShell and nConnect.

In Chord, the participants only have a partial view of the network, and there is no central entity to assist with searches.
When a requester needs to resolve a query, it collaborates with other nodes from the network, asking them if they have the target information.
This poses a privacy challenge, as all the nodes that participate in the routing will know what information is being searched for.
Malicious nodes can exploit this information to punish requesters based on their activity or disrupt their communication.
For example, in Tor, allowing directory nodes to know the query's target can allow them compile statistics about which onion services are being accessed~\cite{eaton2022onion}.

Although privacy was not initially a design objective in Chord,
achieving it is clearly desirable in many (most?)
situations~\cite{srivatsa2004vulnerabilities, wallach2003survey,
  sit2002security}.  In fact, a number of authors already studied the
privacy of structured P2P networks.  However, most existing works
propose anonymity schemes that conceal the requester's
identity~\cite{tarzan2002, shadowwalker2009}.  This is a good option
if the P2P network does not need to provide authentication, but for
authenticated connections, and networks with long term identities,
they break the authentication of the communicating parties.

In this work, we assume a network with authenticated nodes, and we
wish to maintain the authentication while still providing privacy.
Authenticated nodes enable a number of benefits and there are several
existing works proposing different schemes, e.g.,
\cite{butler2009authentication, ingmar2007skademlia,
  prunster2018holistic, aktypi2022themis}.  Existing P2P applications
such as CFS~\cite{frank2001cfs}, NKN~\cite{nkn2018whitepaper}, the
Inter Planetary File System IPFS~\cite{benet2014ipfs} and the Storj
distributed storage platform~\cite{storj2018whitepaper}, all assume
authentication is in place, with CFS and NKN being built on Chord.
Our proposed scheme \alg, achieves search privacy while maintaining
authentication by hiding the content of the requester's queries,
rather than their identity. This allows nodes to have long-term
identities that can be used to initiate queries in the network,
without revealing the content of their search queries.

Since the content of a search query forms the basis of the existing
routing procedure in Chord, hiding that information comes with a
number of challenges. First of all we have to guarantee correctness
(i.e., convergence to the target object) for the new routing
algorithm, while concealing the target from the intermediate nodes.
The intermediate nodes need to know what address the requester is
targeting, to determine how to identify the next hop. We replace the
target with a different one that gets us closer to the real target
without revealing too much information. Getting this right is the crux
of \alg and the details are described in Section~\ref{sec:iris}.

The search algorithm follows an iterative process in which the
requester queries a new node at every step, gradually converging on
the target. Because of this iterative process, the adoption of privacy
notions such as $k$-anonymity to quantify the information leakage,
would result in a different privacy guarantee for each iteration.  We
need a new privacy guarantee that can provide an overall value with
which we can compare strategies and quantify requirements. It needs
sufficient granularity to express the level of information leakage
achieved at every step of the iterative retrieval process. For this we
propose \prv. This notion of privacy will provide the foundation to
argue about the privacy level achieved within a search.

Because Chord is already being used by deployed applications, and the
peer-to-peer nature of the deployments means that there is no central
authority that can mandate a software upgrade, it is critical that our
solution can co-exist with regular Chord nodes, i.e., nodes that do
not run our \alg protocol. The literature contains several proposed
schemes that necessitate significant modification either to the
organization of the nodes~\cite{salsa2006,chill2021} or to the data
structure~\cite{shardis2012}, and we believe that that lack of
compatibility with existing systems is partly to blame for the lack of
adoption. We make sure the design of \alg is backwards-compatible with
the existing search algorithm (and node behavior) for Chord
networks. \alg makes use of the low-level Chord algorithms as building
blocks to achieve this. The requester has the freedom to decide the
level of privacy he wants to achieve, without demanding any deviation
from the vanilla Chord algorithms from the queried nodes.  In this
way, \alg can be used directly in already deployed applications.

The act of concealing the real target of a search introduces a modest
overhead in terms of the number of requests needed to eventually reach
the real target. While some overhead is acceptable as the price of
privacy, we want to make sure \alg is usable in practice. We show
thorough analysis and simulation that the overhead introduced by \alg
is logarithmic in the number of hops, which makes it acceptable. More
importantly, the overhead is proportional to the level of privacy the
requester wants to achieve. That level is tunable by the requester and
zero privacy means zero overhead, so a performance critical
application can pick and choose which searches need (which level of)
privacy.

Our contributions can be summarized as follows:
\begin{itemize}
	\item We propose a new privacy metric, which we call \prv,
	that allows us to quantify the information leakage in structured P2P networks.
	\item We design \alg, a new algorithm that leverages the \ttt{lookup} operation
	inherently built in Chord overlay to allow for query privacy based on the requester's requirements.
	\item We prove the security of our algorithm with respect to the new privacy metric we introduce.
	\item We further confirm empirically through simulations of the communication overhead
	that \alg introduces and the privacy it achieves for different populations of colluding adversarial nodes.
\end{itemize}

\subsubsection*{Paper Structure}
We start with a brief background and notation on Chord structured P2P lookup service in Section~\ref{sec:bg}.
We provide a detailed description of the designing goals and challenges our proposal addresses in Section~\ref{sec:problem}, followed by a formal definition of its system and adversary model in Section~\ref{sec:sysadv}.
We then define \prv, the metric we introduce to quantify privacy in 
networks that follow the Chord lookup service.
We continue by describing \alg, a privacy-preserving algorithm for Chord.
We analyze the privacy guarantees that \alg offers in Section~\ref{sec:security}
and evaluate its performance through simulations in Section~\ref{sec:eval}. 
We compare our approach with prior related works that provide privacy guarantees in P2P architectures in Section~\ref{sec:related}, before closing the paper in Section~\ref{sec:conclusion}.

\section{Background}
\label{sec:bg}

In this section, we provide background on Chord,
the communication scheme on which \alg builds on
enhancing the privacy guarantees it provides.

Chord is a structured P2P network that offers
a decentralized and scalable search service.
It defines how $K$ key-value pairs are stored across $N$ peers
and allows the retrieval of the value associated with a given key
by locating the peer to which this key is assigned.
Both the peers that participate in the network and
the keys that are stored
get an $m$-bit identifier $I$ from an address space.
The address space contains $2^m$ discrete identifiers
from the set $\{0, 1, ..., {2^m}-1\}$ and
is often visualized as a ring.
The peers and the keys are depicted as anchor points on the ring,
sorted in increasing order in a clockwise direction.
A cryptographic hash function $h(\cdot)$ is used to
calculate and to uniformly distribute the identifiers on the address space.
Often---without that being a functional requirement---a
peer's identifier is calculated by applying $h$ on its public key or its IP,
while a key identifier is produced by hashing the key (the data) or its name descriptor.
To distinguish between the identifiers of peers and the identifiers of keys,
we refer to them as \textit{nodes} and
\textit{objects}, respectively.

Each node stores the value of every object
from a range on the address space
in a table referred to as the \textit{object table}.
Each object is assigned to (stored at) the node that is equal to
or follows the object in the address space.
We refer to the node that stores the value of an object
as the \textit{responsible} node for this object.
Due to their uniform distribution
the average distance between the network
nodes and objects is equal to $\dnodes=(2^m-1)/N$ and
$\dobjects=(2^m-1)/K$, respectively.
Thus, on average every node is responsible for $\dnodes/\dobjects$ objects.

Nodes have a partial view of the network,
knowing the communication information of only selected nodes.
Every node saves the details of their \textit{predecessor},
namely the node that comes before them along the address space.
They also save the details of $m$ nodes
that succeed them in the address space
in a table referred to as their \textit{routing table}.
The $j^{th}$ entry of the routing table of a node $N_i$
has the information of the responsible node for
$N_i+2^{j-1}$, where $1\leq j \leq m$.
The node stored in the routing table's first entry
is called the node's \textit{successor}.
This structure ensures that nodes can get the communication information
of every other node in the network, asking no more than $log_2(N)$ other nodes.

\subsection{Modeling Chord}

We model how Chord works, aiming to provide the backbone
on which \alg builds upon and enhances.
We describe only protocols that need to be executed to allow
specific data to be stored in the network and retrieved by a requester.
Thus, we exclude from our model protocols used in network maintenance, \eg \ttt{leave} and \ttt{update}.
We identify four low-level protocols as described below:

\begin{enumerate}
	\item \ttt{bootstrap}$(N_n) \to (N_r, {RT}_r)$:
	Protocol executed between the requester and an existing member
	of the network $N_n$ that returns the address of the requester $N_r$ and the requester's initial routing table ${RT}_r$.
	\item \ttt{lookup}$(N_n, O_k) \to ({N_n}')$:
	Protocol executed between the requester and $N_n$. The protocol
	takes the address of the communication partner node $N_n$, and the
	address of the data object $O_k$, and returns a new node address
	that is closer to the data object. If $N_n = {N_n}'$ the responsible
	node for $O_k$ has been found.
	\item \ttt{fetch}$(N_n,O_k) \to$ data OR nil:
	Protocol to retrieve data with address $O_k$ from node $N_n$.
	\item \ttt{push}$(N_n, O_k, Data) \to$ Ack OR Nack:
	Protocol to upload data with object address $O_k$ to node
	$N_n$.
\end{enumerate}

When a node wants to store or to retrieve data from the network,
it invokes a number of the aforementioned low-level protocols.
We abstract the steps that nodes perform in each case
as two high-level algorithms indicated below:

\begin{enumerate}
	\item \ttt{store}$(RT_h, O_k, Data) \to$ Ack OR Nack:
	Algorithm to store data in the network. As depicted in 
	Algorithm~\ref{alg:store}, it takes three arguments,
	the routing table of the node that holds the data $RT_h$,
	the object $O_k$ of the inserted data
	and the $Data$ itself. It returns a binary status value
	that indicates success or failure.
	\item \ttt{retrieve}$({RT}_r, O_k) \to$ $Data$ OR nil:
	Algorithm to retrieve data from the network. As depicted in
	Algorithm~\ref{alg:retrieve}, it takes the routing table
	of the requester ${RT}_r$, and the object of the requested data.
	The algorithm returns the $Data$ or `nil' to indicate
	that no data can be found at this address.
\end{enumerate}

\begin{algorithm}[tb]
	\caption{Chord's Store Algorithm}
	\label{alg:store}
	\begin{algorithmic}[1]
		\Function{Store}{${RT}_h, O_k, Data$}
		\State $N_n\gets$\Call{SelectClosestNode}{${RT}_h, O_k$}
		\Repeat
		\State ${N_n}'=N_n$
		\State $N_n\gets$\Call{Lookup}{${N_n}', O_k$}
		\Until{$N_n=={N_n}'$}
		\State \textbf{return} \Call{Push}{$N_n, O_k, Data$}
		\EndFunction
	\end{algorithmic}
\end{algorithm}

In both algorithms the executing node needs to identify
the responsible node for the object that it wants to store or retrieve.
Since in Chord there is no centralized entity that can assist with the search,
and nodes do not have a global view of the network,
the node has to ask other nodes, if they
are responsible for the queried object.
The node asks first the node from its routing table
that most closely precedes the target object.
Every queried node checks if the requested object
belongs in the address range
between its own and its successor identifier.
If this is not true, the queried node,
similar to how the initiator picked the first node,
identifies the next queried node.
If the target object is between the queried node and its
successor---there is no node that is closer to the target
than the queried node---the queried node returns
its successor and the recursive execution of the
\ttt{lookup} protocol is terminated.
The initiator then executes with the identified
responsible node the \ttt{push} or \ttt{fetch} protocol
and the algorithm terminates.

\begin{figure}[tb]
	\centering
	\includegraphics[width=1\columnwidth]{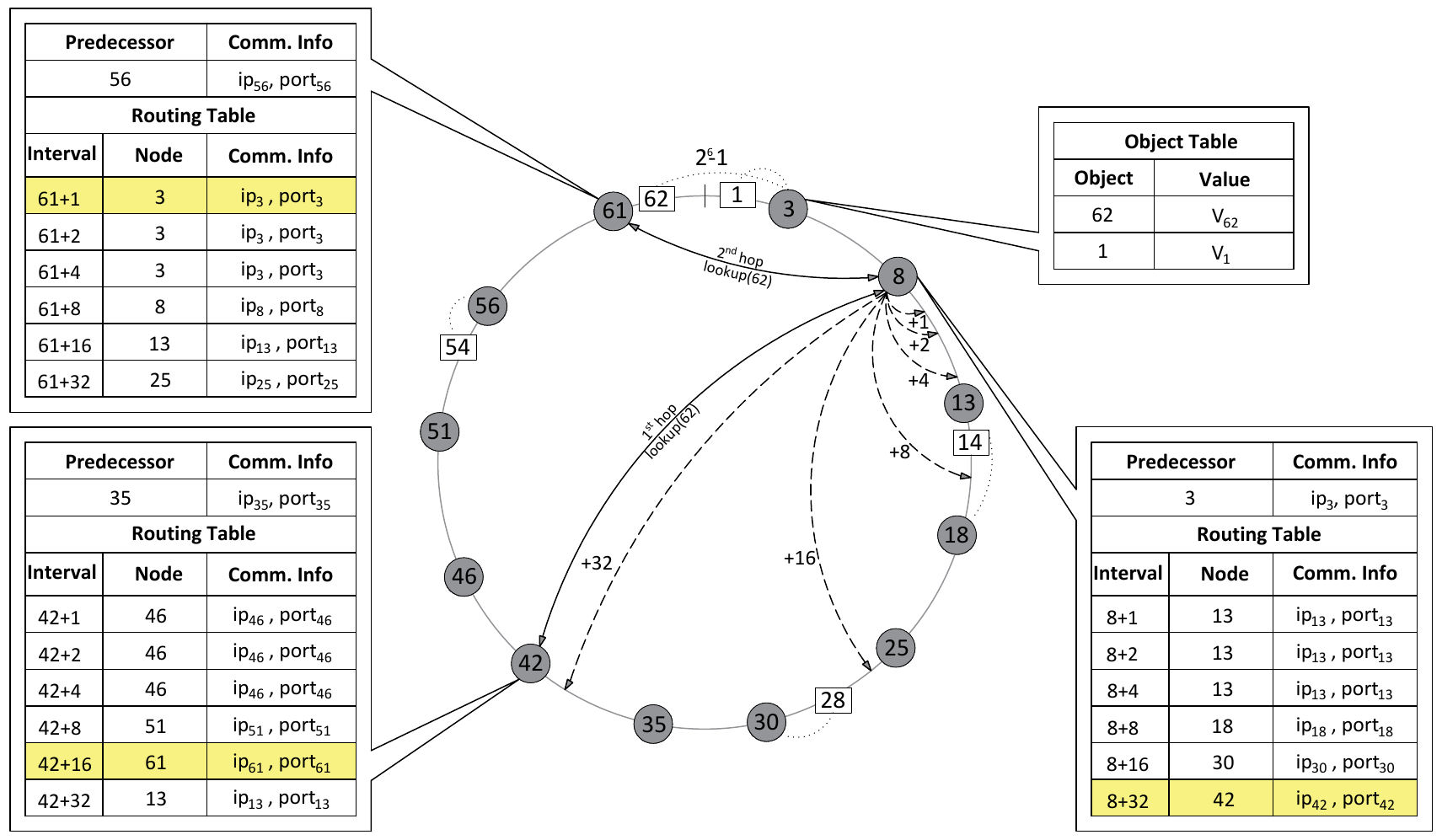}
	\caption{\emph{An example of the Chord's \ttt{retrieve} algorithm.}
			Node $8$ executes \ttt{retrieve} to fetch the data associated with object $62$.
			The participating nodes in the network are depicted as grey circles and
			the registered objects as white squares.}
	\label{fig:ChordlRetrieve}
\end{figure}

In Figure~\ref{fig:ChordlRetrieve}, we can see an example execution
of the \ttt{retrieve} algorithm in a Chord network.
The requester, node $8$ searches
for the responsible node of object $62$.
Initially, the requester checks its routing table to identify the
node that most closely precedes object $62$
and selects node $42$ with which it executes the \ttt{lookup} protocol first.
As object $62$ is not between node $42$ and its successor, \ie node $46$,
node $42$ relays the requester to node $61$.
Next, node $8$ executes \ttt{lookup} with node $61$.
For node $61$, the queried object $62$
is between its own and its successor identifier, \ie node $3$;
thus, node $3$ is the responsible node for object $62$.
Node $61$ responds to node $8$ by specifying node $3$.
Node $8$ executes with node $3$ the \ttt{fetch} protocol
and the \ttt{retrieve} algorithm is terminated.

\begin{algorithm}[tb]
	\caption{Chord's Retrieve Algorithm}
	\label{alg:retrieve}
	\begin{algorithmic}[1]
	\Function{Retrieve}{${RT}_r, O_k$}
		\State $N_n\gets$\Call{SelectClosestNode}{${RT}_r, O_k$}
		\Repeat
			\State ${N_n}'=N_n$
			\State $N_n\gets$\Call{Lookup}{${N_n}', O_k$}
		\Until{$N_n=={N_n}'$}
		\State \textbf{return} \Call{Fetch}{$N_n, O_k$}
	\EndFunction
	\end{algorithmic}
\end{algorithm}

\section{Problem Statement and Design Goals}
\label{sec:problem}
This section explains the challenges of developing
a private query mechanism in Chord, followed by
an outline of the design goals such a mechanism must achieve.

\subsection{Problem Statement}

The adoption of Chord P2P networks in real-world applications such as
CFS and NKN, underscores the critical need to provide robust privacy
guarantees in these networks.

When a node is asked for the location of a target, as part of the
search process, the target is an address that does not by itself
reveal much information. It is essentially a hash of the
content. However, the node is free to request that same target
himself, and obtain the corresponding data. This allows any node in
the network to monitor and track the data that is being searched for
by others. Because each node in an authenticated Chord network has a
long-term identity, it is possible to build a profile of each identity
by tracking network activity, and reveal additional information over
time. This information, when exploited by actors such as surveillance
agencies, advertisement companies, or states that apply censorship,
can have severe implications.

As described in Section~\ref{sec:bg}, the Chord \ttt{lookup} protocol
requires that the requester reveals the target object to every queried
node. The nodes decide where to forward a message based on the target
address. In every hop, the distance to this address gets smaller,
guaranteeing convergence.  A trivial solution that replaces the target
with a random identifier cannot guarantee convergence (in a reasonable
amount of time) and thus, cannot be applied. Picking a fixed address
calculated based on a predefined offset may not reveal the target
directly, but it still allows for easy discovery if the offset is
known or can be guessed. 

Even assuming we can hide the target address in the request, some
information can be obtained by looking at the relative position
(address) of the requester. The Chord search algorithm dictates that
requester selects the node from its routing table that most closely
precedes the target. Knowing the address of the requester, and the
structure of the nodes in a normal routing table, a node can narrow
down where in the address space the target object is likely to be.

\subsection{Design Goals}
In the design of a privacy preserving lookup algorithm for \alg, we
consider the following objectives:

\begin{enumerate}
\item \emph{Correctness}: convergence to the node responsible for the
  target must be guaranteed in a reasonable amount of time (hops).
  Ideally the trade-off between the level of privacy and the
  convergence speed should be determined by the requester, on a
  per-object basis.
\item \emph{Query privacy}: given an authenticated requester, the
  inference that any malicious queried node makes regarding the target
  of the query should not violate the level of \prv chosen by the
  requester for that query.
\item \emph{Interoperability}: hiding the target object should
  leverage already deployed infrastructure without any change in the
  network organization or in the communication of the queried
  nodes---\alg should work, and be secure, even if the requester is
  the only node using it.
\end{enumerate}

Finally there is the question of whether to conceal the target address
from the final node in the search process. The one that is responsible
for the target. We have chosen not to incorporate that into \alg, if
such a property is required one can use a number of existing options
to accomplish that, e.g., Private Information Retrieval, or a more
naive solution where the nodes sends all the objects it controls. We
consider this to be an orthogonal problem to the one of providing
privacy from the nodes along the search path, and we will not address
that further in this paper.

\section{System and Adversary Model}
\label{sec:sysadv}

We assume a set of $N$ nodes and $K$ named data objects, organized in
an authenticated Chord peer-to-peer network as described in
Section~\ref{sec:bg}. Nodes can communicate directly as long as they
have each other's communication details (IP address, etc.).
Communication is done on top of an authenticated channel, e.g.,
\cite{palomar2006authentication, butler2009authentication,
  aktypi2022themis}, and as a consequence nodes cannot lie about
their long term identity or network address. The requester's goal is
to be able to search for arbitrary data objects without revealing the
nature of the data to any intermediary nodes as part of the search
process.

We consider an internal attacker who participates in the routing
process by controlling a fraction $f$ of the nodes of the network.
The attacker can act through all the nodes under his control by
initiating requests or responding to incoming connections, but cannot
identify or listen to connections among the remaining honest nodes.
The attacker is active, deviating from the Chord algorithm at will,
\eg redirecting the requester to another malicious node, claiming
responsibility for an object or initiating requests to enumerate the
registered nodes and objects. However, the attacker cannot break the
underlying authentication scheme used in the routing algorithm; thus
cannot lie about the address they control. The adversary knows \alg
and to make the adversary as powerful as possible we give him full
knowledge of the $\alpha$ and $\delta$ parameters the requester
chooses. This would not normally be known to an attacker but we choose
to provide them to the attacker in our model, to account for the
possibility that these parameters could be guessable in a practical
scenario. They are chosen by the user after all, so maybe some common
choices (or software defaults) emerges.

The attacker's goal is to discover the target object of a
query. Specifically the attacker must know the target object with a
probability higher than that allowed by the \mbox{\prv} notion described in
Section~\ref{sec:privacynotion}, for parameters $\alpha$ and $\delta$
chosen by the requester.

\begin{figure}[tb] 
	\centering
	\includegraphics[width=0.75\columnwidth]{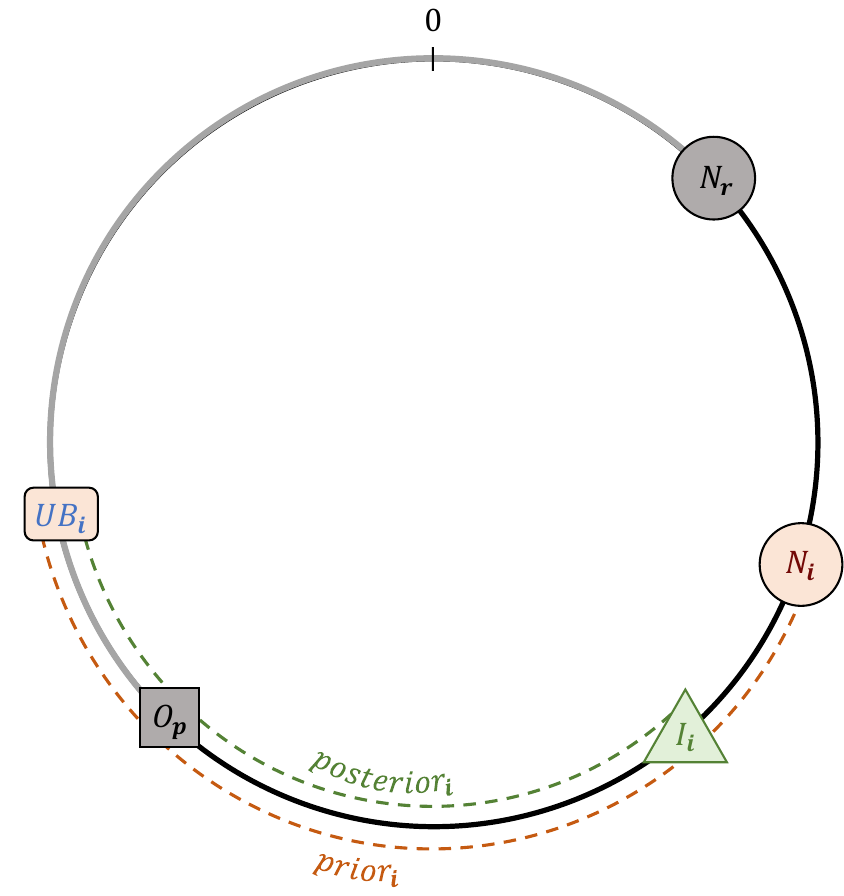}
	\caption{\emph{The privacy metric.}
		The orange dashed line indicates the ${prior}_i$ range of $N_i$ against $N_r$. The green dashed line shows the ${posterior}_i$ range of $N_i$ after knowing $I_i$.
		Both ranges are computed based on ${UB}_i$, \ie the upper bound of node $N_i$’s estimate regarding the range in which belongs the actual target of node $N_r$.}
		\label{fig:notion}
\end{figure}

\section{Alpha-Delta Privacy}
\label{sec:privacynotion}

The development of a privacy preserving algorithm necessitates the
need for a way to measure the privacy guarantees it provides.  Due to
the iterative process of the Chord \ttt{retrieve} algorithm, during
which different nodes with different distances to the target are
queried every time, privacy notions such as $k$-anonymity are not
suitable as they only allow us to argue about the privacy achieved at
every step of the retrieval process and not the retrieval algorithm as
a whole.  To overcome that, before designing our privacy preserving
retrieval algorithm, we introduce a new \emph{privacy notion} that
offers the granularity to argue about the privacy achieved at the
level of a completed query, \ie that simultaneously applies to every
step that a search incorporates.

To solve this problem we define \prv\ to measure the privacy level of
an \alg{}-request.  This notion is parameterized by two values
$\alpha$ and $\delta$ which are chosen freely by the requester. They
can be different for each new search the requester performs, and could
in theory even be changed between iterations of a single run of
\alg. Despite this, our adversary model requires that the attacker
knows both of these values. All our results are presented with this in
mind and can thus be considered the worst-case privacy for the
requester.

The parameter $\delta$ serves a similar role to $k$ in $k$-anonymity
as it describes the size of the initial anonymity-set in which the
real target object must reside. $\alpha$ describes how quickly we
progress towards the target object, and thus the decay in privacy per
iteration. In order to understand how this works it is necessary to
introduce a few details about the search process.

In each iteration of \alg a potential attacker is queried for the
address of a target node. We explain this process in detail in
Section~\ref{sec:iris} but for now it is sufficient to know that this
query reduces the size of the address range where the actual target
can be by a certain amount. This gives rise to two address ranges, a
${prior}_i$ range that an attacker could deduce from knowledge of
$\delta$ and previous search iterations performed with colluding
attackers, and a ${posterior}_i$ range that incorporates the knowledge
gained from the ongoing request. These two ranges are visualized in
Figure~\ref{fig:notion}, note that ${posterior}_i$ is always smaller
than ${prior}_i$. With this we can describe $\alpha$ as the minimum
allowable ratio between the posterior and prior knowledge of a
requester in any iteration.

\begin{align*}
	\alpha \le \textrm{min}_i\left(\frac{{posterior}_i}{{prior}_i}\right) \forall i
	\label{eq:privacy_notion}
\end{align*}

An equivalent way to think about $\alpha$ is as (one minus) the
maximum allowable gain in knowledge by any intermediate node. We can
now state the definition of \prv.

\begin{definition}[\prv]
	A search algorithm is \prvt if the following two conditions hold:
	(1) ${prior}_0 \ge \delta$ for the first queried node $N_0$; and
	(2) ${posterior}_i/{prior}_i \ge \alpha$ for every iteration $i > 0$
\end{definition}

Choices for $\delta$ are values in the interval $[0,2^m-1]$ where $2^m$
is the size of the address space.
Similarly choosing $\alpha \in [0,1)$ allows a requester to tune the trade off
between privacy and performance. The closer $\alpha$ is to $1$ the less
additional information $N_i$ gains about the intended target.

\begin{table}[tp]
	\renewcommand{\arraystretch}{1.2}
	\centering
	\caption{List of symbols and notation used in this paper}
	\label{table:tasks}
    \begin{tabular}{@{}ll@{}}
		\toprule
		$2^m$ & size of the address space \\
		$N$ & number of participating nodes in the network \\
		$K$ & number of registered objects in the network \\
		$N_r$ & identifier of the requester \\
		$RT_r$ & routing table of the requester \\
		$O_p$ & identifier of the target object \\
		$N_t$ & identifier of the node responsible for $O_p$ \\
		$\alpha, \delta$ & privacy parameters explained in Section~\ref{sec:privacynotion}\\
		$S$ & starting address of the search $S = O_p - \delta$ \\
		$N_i$ & identifier of the node being queried \\
		$N_{i+1}$ & identifier of the next node to be queried \\
		$R_i$ & reference point selected against node $N_i$: $R_i \in_R [N_i, O_p)$ \\
		$I_i$ & identifier for which $N_i$ is queried: $I_i=R_i+(N_i-R_i)\cdot \alpha$ \\
		$d_i$ & distance between $N_i$ and $O_p$ \\
		${U\!B}_i$ & upper bound of the target range that node $N_i$ can estimate \\
		$prior_i$ & target range $N_i$ can estimate by knowing $\delta$\\
		$posterior_i$ & target range $N_i$ can estimate by knowing $\delta$ and $I_i$\\
		$f$ & fraction of colluding adversaries in the network\\
		\bottomrule
	\end{tabular}
\end{table}

\section{\alg}
\label{sec:iris}
In this section, we describe \alg,
the mechanism we develop to allow for
\prvt queries in Chord.
Table \ref{table:tasks} defines the symbols and notation we use.
We start by outlining the core idea behind its design.
We then provide a detailed description
with an execution example.

\subsection{Overview}
\label{sec:overview}

The ordered address space that is leveraged by the nodes in Chord provides a numerical basis
to position nodes and calculate the distance between them.
In every hop, the requester, by asking for the target object,
finds nodes that have a smaller distance than the requester to the target.
Yet, finding nodes that satisfy this condition
can also be achieved if instead of the target object
the requester queries for another identifier
that is between the requester and the target object.
Due to the ordered address space,
getting closer to this intermediate identifier
allows simultaneously the requester to get closer to the target object.

We built on this observation to develop \alg.
In \alg, the requester, rather than asking the queried node
for the target object, asks for an intermediate identifier.
In this way, the requester gets closer
to the target without however revealing the target.
The requester iterates this process
asking every time for another intermediate
identifier so as to find the responsible node for the target object.
In the section below we define how this process is done.
We make \alg such as achieving \prv
when using Chord.

\subsection{Mechanism Description}
\label{sec:iris_description}

\alg replaces the regular \ttt{retrieve} algorithm from Chord.
It takes two additional parameters, $\alpha$ and $\delta$
that determine the level of privacy to use for the request:
\ttt{iris}$({RT}_r, O_p, \alpha, \delta) \to$ $Data$ OR nil.

The $\delta$ parameter allows the requester
to control the size of the address range to which the target belongs,
which the queried node can estimate.
In Chord's \ttt{retrieve} algorithm,
the requester asks first the node in its routing table
that is closest and does not succeed the target address.
This greedy heuristic gives to the queried nodes
an estimate regarding the target of the request---the target
does not succeed the requester's successor,
which is deterministically defined, that comes after the queried node.
Because nodes have a more dense view of the address space
closer to them, this estimate becomes more accurate
the closer the queried node is to the requester.
To overcome this leakage \alg modifies Chord's \ttt{retrieve} protocol
node selection by changing how the requester picks the first node to query.
The $\delta$ parameter is used by the requester
to calculate an address $S$ that precedes the target object $O_p$ by $\delta$.
The requester selects the first node to query
to be its successor that most closely succeeds $S$ but precedes $O_p$.
If none of its successors belongs in this interval
the first node to be queried is the requester's successor
that most closely precedes $S$.
With this selection process every queried node $N_i$,
regardless of how close to the requester is,
assuming that the node knows $\delta$,
can only deduce that the target of the request is
one out of $\delta$ addresses from the address space that succeed $N_i$.

The $\alpha$ parameter controls how fast
the request converges to the target.
The requester hides the target of the query
from the intermediate nodes by substituting
$O_p$ with another address $I_i$,
which succeeds the queried node but precedes the target.
To pick $I_i$ the requester firstly selects a reference point $R_i$
by selecting uniformly at random an address in the interval $[N_i, O_p)$.
By bounding the range selection of the randomly picked points
to the actual target of the query,
\alg guarantees correctness with the least possible number of steps,
at the cost of allowing colluding attackers
that get closer to the target
to have a better guess regarding the target,
\ie being able to calculate a $prior$ range of smaller size.
The requester then calculates $I_i$
as the linear interpolation between the address of the queried node
and the reference point, based on the $\alpha$ parameter, \ie
$I_i=R_i+(N_i-R_i)\cdot \alpha$.
The reference point provides privacy to the requester
against a colluding adversary.
Calculating $I_i$ as the linear interpolation between the address of the queried node
and the target, based on the $\alpha$ parameter, \ie
$I_i=O_p+(N_i-O_p)\cdot \alpha$, would provide no privacy against
our strong adversary model.
By querying $N_i$ for $I_i$ the requester
in every hop converges by a rate $\alpha$ to the reference point.
As $R_i$ comes after the queried node and precedes the target,
the requester by converging to $R_i$ converges simultaneously to $O_p$.

\begin{algorithm}[tb]
		\caption{\alg's Retrieve Algorithm}
		\label{alg:iris}
	\begin{algorithmic}[1]
		\Function{Iris}{${RT}_r, O_p, \alpha, \delta$}
		\State $N_i\gets$\Call{SelectStartNode}{${RT}_r, O_p, \delta$}
		\Repeat
		\State ${N_i}'=N_i$
		\State $R_i\gets$\Call{RandomAddressBetween}{$N_i, O_p$}
		\State $I_i\gets$\Call{lerp}{$R_i, N_i, \alpha$}
		\State $N_i\gets$\Call{Lookup}{${N_i}', I_i$}
		\Until{$N_i=={N_i}'$}
		\State \textbf{return} \Call{Fetch}{$N_i, O_p$}
		\EndFunction
	\end{algorithmic}
\end{algorithm}

Algorithm~\ref{alg:iris} depicts \alg's pseudo-code.
After calculating $I_i$, the requester
asks $N_i$ for $I_i$ leveraging the Chord \ttt{lookup}.
The queried node following the Chord \ttt{lookup} algorithm
replies to the requester by indicating
either another node closer to $I_i$ or its responsible node.
If the node to which the requester is relayed
still precedes the target identifier, the requester repeats the process.
The requester is free to renew the value of $\alpha$
between iterations.

The requester stops querying nodes
when being relayed to a node that succeeds $O_p$.
In this case, due to how the ownership of objects is defined in Chord,
the responsible node of the target is simultaneously
responsible for the queried identifier.
The requester then precedes by executing the \ttt{fetch} protocol
with the node responsible for $O_p$, getting back either the $Data$ or $nil$
if this object does not exist, and the \ttt{iris} algorithm terminates.

\begin{figure*}[tb] 
	\centering
	\includegraphics[width=\textwidth]{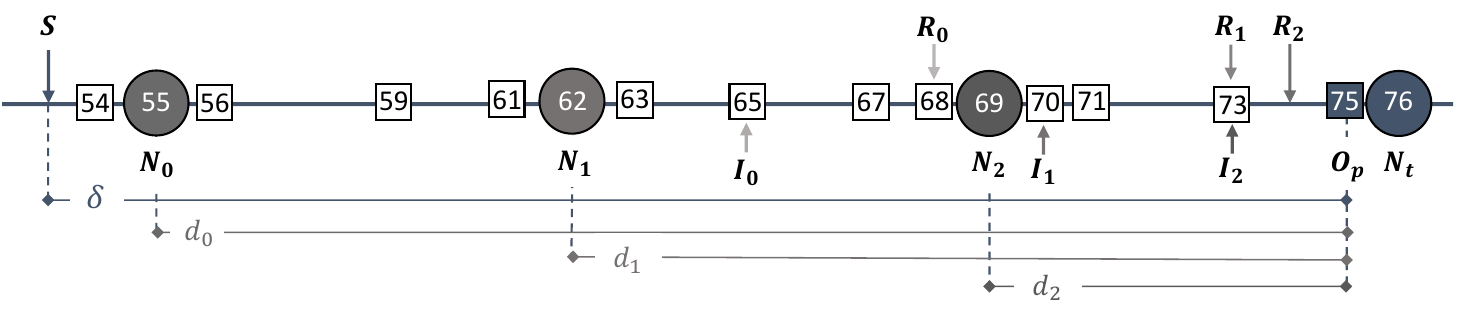}
	\caption{\emph{\alg's application example.}
		The requester targeting object $O_p=75$ selects $\delta=22$ and $\alpha=0.25$.
		queries back to back nodes for identifiers chosen in the interval $[53,75)$.
		In every iteration the interval degrades, converging at the end to node $N_t=76$.}
	\label{fig:IRISexample}
\end{figure*}

\textbf{Application Example:} An example execution of \alg
is depicted in Figure~\ref{fig:IRISexample}.
Let's assume a requester node
$\reqexample$ wants to find the node that stores the values of
a service by the name $'secret'$.
To achieve that, $\reqexample$ needs to identify
the responsible node of the targeted object $O_p=h('secret')=75$.
To avoid revealing the targeted object to the network,
node $\reqexample$ executes \alg.
First, $\reqexample$ tunes the $\delta$ parameter to be equal to $22$
and calculates $S=53$ by abstracting $22$ from $75$.
Node $\reqexample$ after consulting its routing table
selects $N_0=55$, as this is the first
of its successors belonging in the interval $[53,75]$.
Node $\reqexample$ then selects parameter $\alpha=0.25$
to control the rate of convergence of the query.
After picking randomly $R_0=68$ in the interval $[55,75)$,
node $\reqexample$ calculates $I_0$ based on $|I_0-68|=0.25 \cdot|55-68|$,
thus, queries node $55$ for the identifier $65$.
Node $55$ relays the requester to node $62$.
The requester then checks if node $62$
comes after the targeted object $75$.
As this is not the case, it continues by picking
a new random identifier $R_1=73$ from the range
$[62, 75)$ and calculating a new queried identifier $I_1=70$.
Node $\reqexample$ queries node $62$ for $70$
having, as a result, to be relayed to node $69$.
Finally, the node $69$ when queried for $73$---calculated based on the reference point
$74$---relays node $\reqexample$ to node $76$
that comes after the target object $75$.
Node $\reqexample$ executes the \ttt{fetch} protocol with node $76$
asking for object $75$
to retrieve the stored mapped data.

\section{Security Analysis}
\label{sec:security}

In this section we analyze \alg's \textit{correctness}
and we prove that \alg is an \prvt algorithm.
We further analyze theoretically the advantage that
\alg gives to a powerful adversary.

\subsection{Correctness}
\label{sec:correctness}

We start the security analysis of \alg by formally proving its correctness,
\ie guaranteeing that the requester upon executing the algorithm
succeeds in identifying the node that stores the searched value.

	Let $N_t$ be the first successor of the requester's target object $O_p$,
	\ie $N_t$ is the responsible node of $O_p$.
	Consider the $i$-th iteration of the algorithm
	where the requester executes the Chord \ttt{lookup}
	with node $N_i$ specifying the address $I_i$,
	getting back $N_{i+1}$ that is the next node to query.
	Recall from Section~\ref{sec:iris_description} that the requester
	queries node $N_{i+1}$ for another address $I_{i+1}$;
	thus, $N_{i+1}$ is not necessarily the predecessor of $I_i$,
	yet, due to how the \ttt{lookup} progresses $N_{i+1}$ is closer than $N_{i}$  to $I_{i}$.
	
	Let $d_i$ be the distance between $N_i$ and $O_p$,
	and $R_i$ be a uniformly randomly picked address in $[N_i, O_p)$.
	On average, $R_i$ is picked in the middle of the interval, thus,
	$|R_i-N_i|=|O_p-R_i|=d_i/2$.
	The queried address $I_i$ is calculated based on
	$R_i$ and the parameter $\alpha$ that the requester selects.
	More precisely, $I_i$ is selected such as
	$|R_i-I_i|=\alpha\cdot|R_i-N_i|$; thus, $|R_i-I_i|=\alpha\cdot d_i/2$.
	Assuming that the distance between $N_{i+1}$ and $I_i$ is zero
	we have that $|R_i-N_{i+1}|=\alpha\cdot d_i/2$.
	From Figure~\ref{fig:IRISexample}, we observe that the distance
	the requester has to $O_p$ at the $(i+1)$-th iteration
	is the sum of the distance the queried node has to $R_i$
	plus the distance of $R_i$ to $O_p$.
	Considering the above calculations,
	this distance is equal to
	$d_{i+1} = \alpha \cdot d_i/2+d_i/2$.
	By referencing every step to the initial distance
	the requester has to the target object $d_0$,
	we have that in the $n$-th iteration
	the distance of the requester to $O_p$ is given by ~\Cref{eq:distance}.
	Because $\alpha \in [0,1)$, we have that $\lim_{n \to \infty} d_n = 0 $,
	thus, the algorithm converges on the target.

		\begin{equation} 
			d_n = d_0 \cdot \left(\frac{\alpha+1}{2}\right)^n
			\label{eq:distance}
		\end{equation}

	Nodes are responsible for the identifiers that fall between
	their predecessor and their own node identifier;
	thus, the responsible node for $O_p$,
	follows $O_p$ on the address space
	and there is no other node placed between them.
	\alg has the selection interval of the selected queried identifiers
	not to exceed the targeted object.
	Querying identifiers that only precede the targeted object
	guarantees that if a randomly picked
	object has its responsible node
	succeeding $O_p$,
	this node is also responsible for $O_p$.
	The \ttt{iris} algorithm is terminated to the predecessor
	of the node that is responsible for the queried identifier.
	Thus, \alg terminates when the distance the requester has
	to the target object becomes equal to the average distance
	the nodes have on the address space $\dnodes$.
	Setting $d_n=\dnodes$,
	the number of iterations
	the requester needs on average
	to identify $N_t$ while executing \alg is:

	\begingroup
	\fontsize{9.5pt}{5pt}\selectfont
	\begin{align}
		d_n=\dnodes
		&\Rightarrow (\alpha+1)^n \cdot \frac{d_0}{2^n} = \dnodes \nonumber\\
		&\Rightarrow n = \frac{log_{\alpha} d_0-log_{\alpha} \dnodes}{log_{\alpha} 2 - log_{\alpha} (\alpha+1)}
		\label{eq:hops}
	\end{align}
	\endgroup

\subsubsection{Secure Routing}

Based on the \ttt{lookup} protocol, the requester is relayed to
every other but the first node by the previous queried node.
This can be leveraged by a colluding attacker who can relay the requester
to a malicious node that---given that it succeeds the target---will
be accepted by the requester as the responsible node,
thus, learn the target.

Let's assume that the requester is relayed to $N_{i+1}$ from $N_i$.
The requester to conclude if $N_{i+1}$ is the responsible node for $O_p$,
can use bound checking~\cite{secureRouting}.
Assuming $N$ active nodes,
the distances between consecutive active nodes 
can be modeled as approximately independent exponential random variables with mean
equal to $\dnodes=(2^m-1)/N$.
Given $f \cdot N$ colluding nodes,
the distances between consecutive colluding nodes 
can also be modeled as approximately independent exponential random variables with mean
equal to $d_a=(2^m-1)/(f \cdot N)$.
Let $\mathcal F_1$ and $\mathcal F_2$ be the distributions of
active and colluding nodes, respectively.
The requester can identify if $N_{i+1}$ is the responsible node for $O_p$
by determining if $d_x$, that is the distance between $N_{i+1}$ and $N_i$,
is drawn from distribution $\mathcal F_1$ and not $\mathcal F_2$.
The requester does not know $N$ but based on
its routing table and its distance to its predecessor,
can make an estimation $d_r$
concerning the address range for which each node is responsible.
For $N_{i+1}$ to be $N_t$
we need to have $d_x \ge T$
where $T=\gamma \cdot d_r$ and $\gamma \in (1, 1/f)$.
Based on our threat model the attacker does not control the responsible node, thus $d_x > T$.
According to~\cite{secureRouting}, to obtain the minimum
false positives and false negatives, $\gamma$ must be equal to $\gamma=1/f$.

\begin{figure*}[!t] 
	\centering
	\includegraphics[width=\textwidth]{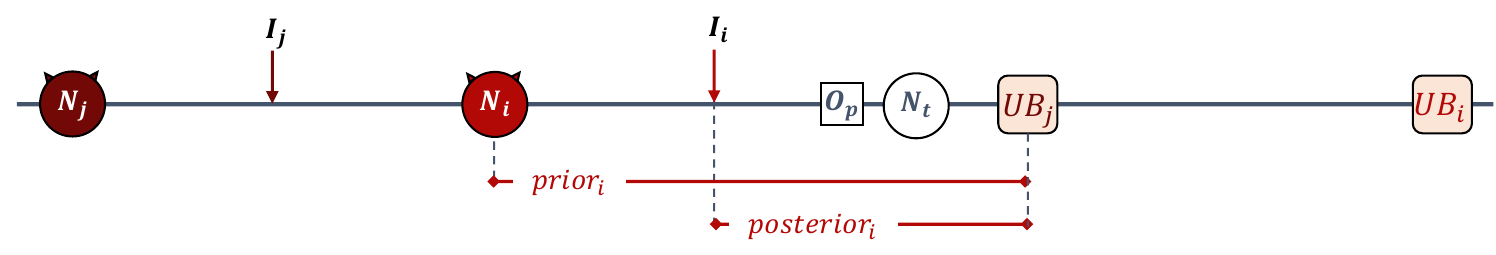}
	\caption{\emph{A colluding adversary.}
	Assuming that $N_j$ is the first asked colluding adversary, every other colluding node that the requester queries can use $UB_j$ instead of $UB_i$ in their calculation to infer the target.}
	\label{fig:anonymity}
\end{figure*}

\subsection{Query Privacy}
\label{sec:query-privacy}

Here we prove that \alg is an \prvt algorithm
following the definition introduced in Section~\ref{sec:privacynotion}.
We start by analyzing
the query privacy guarantees that \alg provides against
an adversary that has no more than one node under control
that can, however, be any of the queried nodes.
We then continue by considering
a more powerful adversary that controls
multiple nodes in the network.
Based on our system and adversary model,
in our analysis we assume that
nodes are authenticated, i.e.,
they cannot lie about the address they control.

\subsubsection{Lone Adversary}

Let us now consider a lone adversary that controls only node $N_i$.
Recall from Section~\ref{sec:iris_description}
the way the requester selects the queried nodes,
\ie $N_0$ is the requester's successor immediate
after or before address $S$ that precedes the target by $\delta$.
Based on this selection process,
$N_i$ can deduce the following about the requester's target.
If one of the requester's successors belongs in $[N_i, N_i+\delta]$
then the requester searches something that belongs in $[N_i, UB_i]$
where $UB_i=N_i+\delta$.
If there are no successors of the requester in $[N_i, N_i+\delta]$
then the requester searches something that belongs in $[N_i, UB_i]$
where $UB_i=N_i+\delta+x$,
denoting as $x$ the larger than $\delta$ distance a queried node can have from the actual target.
Based on our adversary model we assume that the $\delta$ parameter
is known to the attacker.
Thus, the worst scenario is the prior knowledge of $N_i$
to be equal to $|UB_i-N_i|=\delta$.

Node $N_i$ is queried by
the requester for the identifier $I_i$;
thus, the posterior knowledge of $N_i$
is $|UB_i-I_i|$.
$I_i$ is picked based on
the reference point $R_i$ that succeeds $I_i$ but
precedes $O_p$ thus $UB_i$ on the address space.
Thus, the following holds
$|UB_i-R_i| + |R_i-I_i|=|UB_i-I_i|$ and
$|UB_i-R_i| + |R_i-N_i|=|UB_i-N_i|$.
By definition $|R_i-I_i|=\alpha \cdot |R_i-N_i|$,
thus, we have:

	\begingroup
	\fontsize{8.5pt}{5pt}\selectfont
	\begin{align}
			\frac{{posterior}_i}{{prior}_i}
			&=\frac{|UB_i-I_i|}{|UB_i-N_i|}
			= \frac{|UB_i-R_i| + |R_i-I_i|}{|UB_i-R_i| + |R_i-N_i|}\nonumber\\
			&=\frac{|UB_i-R_i| + \alpha\cdot|R_i-N_i|}{|UB_i-R_i| + |R_i-N_i|}\nonumber\\
			&=\alpha \cdot \frac{\frac{|UB_i-R_i| }{\alpha}+|R_i-N_i|}{|UB_i-R_i| + |R_i-N_i|}\nonumber\\
			&=\alpha \cdot c_a
		\label{eq:privacy_ratio_lone}
	\end{align}
	\endgroup

In \Cref{eq:privacy_ratio_lone}, as $c_a$ has in its numerator
the parameter $\alpha\in[0,1)$ as denominator
we can conclude that $c_a>1$,
thus, the ratio between the posterior and prior
knowledge of the adversary is greater than $\alpha$.
Hence, we can conclude that \alg is an
\prvt algorithm against a lone adversary.

\subsubsection{Colluding Adversary}

Let us now consider the case of a colluding adversary
that controls a fraction $f$ of the nodes in the network
with $N_j$ and $N_i$ being two consecutive adversarial nodes,
both queried by the requester when searching for $O_p$.
The worst case scenario is for
the attacker correctly to assume that
the two different queries serve the same search.
Due to the random reference point
in the calculation of the queried address
at step 6 in Algorithm~\ref{alg:iris},
the attacker cannot calculate $O_p$.
However, from Figure~\ref{fig:anonymity},
we observe that node $N_i$ can use as an upper bound
$UB_j$ instead of $UB_i$.
Thus, the prior knowledge of $N_i$ is equal to $|UB_j-N_i|$.
Now considering that $UB_j=N_j+\delta$, we have:

\begin{equation}
	prior_i=\delta-|N_i-N_j|
	\label{eq:privacy_ratio_colluding}
\end{equation}

A colluding attacker with average distance between
colluding nodes bigger than $\delta$ is only queried once,
hence, this case is equal to a lone attacker
from a security perspective.
From \Cref{eq:privacy_ratio_colluding},
we observe that
for a colluding attacker for whom the nodes under control
have average distance $d_a$,
the minimum distance the first adversarial
node has to the estimate upper bound is equal to $\delta$.
However, assuming that $t$ colluding nodes are queried throughout
the \ttt{iris} execution, at the end
the minimum distance the adversarial
node has to the estimate upper bound is equal to
$\delta'=\delta-t \cdot d_a$.

Regarding the prior and posterior knowledge ratio of $N_i$,
if in \Cref{eq:privacy_ratio_lone} we replace $UB_i$ with $UB_j$
we have $posterior_i/prior_i=\alpha \cdot c_b$.
As $UB_i$ succeeds $UB_j$, we have that
$|UB_i-R_i|>|UB_j-R_i|$, thus, $c_a>c_b$.
Hence, \alg achieves a lower
ratio between the posterior and prior knowledge
against a colluding adversary
compared to a lone adversary,
yet still lower bounded by $\alpha$.
From the above we can conclude that
\alg is an \prvt algorithm against a colluding adversary.

\subsection{Attacker Advantage}
\label{sec:attacker-advantage}

Let us now consider the advantage that \alg
gives to the attacker, what the attacker can deduce
regarding the target of the requester
based on the information available to the attacker.
We consider the worst case scenario, assuming
the attacker knows the $\alpha$ and the $\delta$
parameters the requester has chosen.
Following \alg, the requester chooses $I_i$
based on $\alpha$ and the randomly picked address $R_i$.
Assuming the attacker knows $\alpha$, when queried for
an identifier $I_i$, the attacker can calculate
the $R_i$ the requester picked.
Based on this deduction,
in our analysis we examine the probability \alg gives
for the target address to have a specific value $o$
given a randomly picked value $x$ by the requester.

We have assumed that the attacker knows the $\delta$ parameter.
This prior knowledge, as in previous section explained,
can be used by the attacker to calculate an
upper bound for the address of the target, \ie the target
will be an address that is no further away than $\delta$ addresses
from the attacker's address.
However, this is only true for the attackers that succeed $S$ address,
where  $S=O_p-\delta$.
Any attacker preceding $S$, even with the knowledge of $\delta$,
cannot calculate a correct upper bound, \ie the target will succeed
the attacker's address by more than $\delta$ addresses.
The attacker has no way to conclude where on the address space
is placed in reference to the address $S$, thus, the best thing
the attacker can do is to guess that succeeds $S$
even if this might not be the case.

\begin{figure*}[tb]
	\centering
	\subfigure[$P(O_p=o|R_i=x)$]{
		\includegraphics[width=0.95\columnwidth]{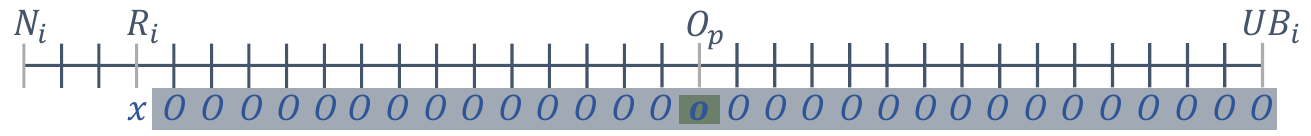}
		\label{fig:Cal_P(O=o|R=x)}}
	\subfigure[$P(O_p<=o|R_i=x)$]{
			\includegraphics[width=0.95\columnwidth]{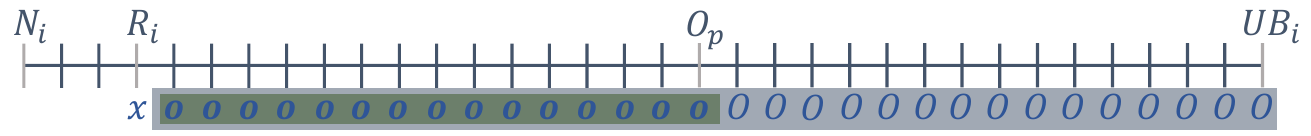}
			\label{fig:Cal_P(O<=o|R=x)}}
	\subfigure[$P(O_p=o|R_i<=x)$]{
		\includegraphics[width=0.95\columnwidth]{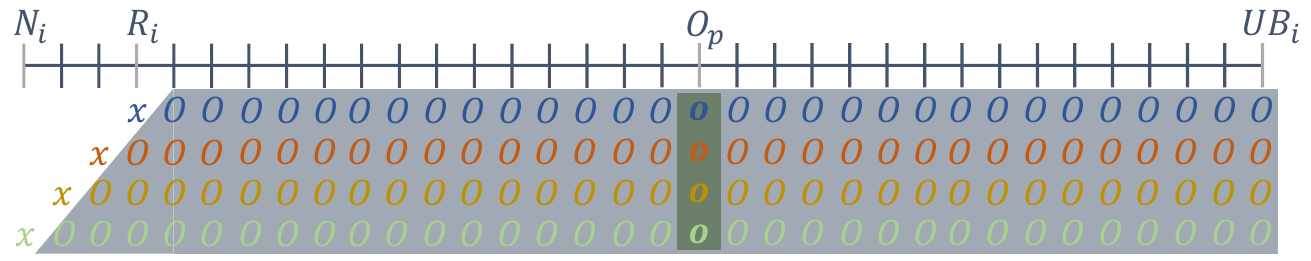}
		\label{fig:Cal_P(O=o|R<=x)}}
	\subfigure[$P(O_p<=o|R_i<=x)$]{
			\includegraphics[width=0.95\columnwidth]{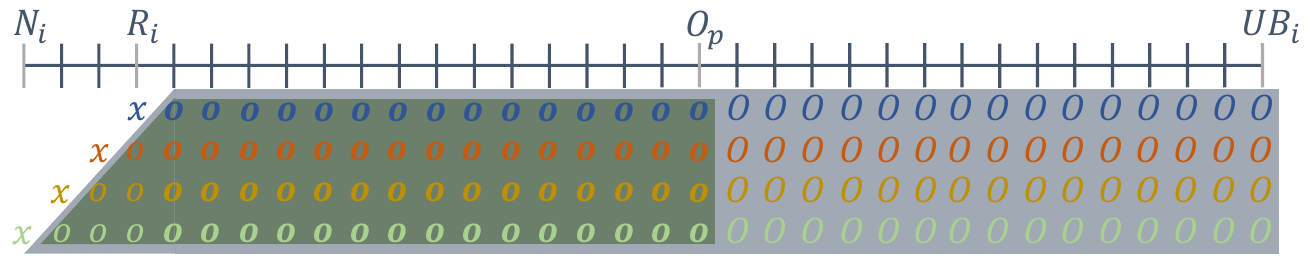}
			\label{fig:Cal_P(O<=o|R<=x)}}
	\caption{\emph{Probability Calculation:} We mark with $O$ the addresses that $O_p$ can obtain
		and with $o$ and $x$ the explicit value(s) of $O_p$ and $R_i$, we examine.}
	\label{fig:Cal_P}
\end{figure*}

Without loss of generality we can position
the attacker at the start of the address space $N_i=0$.
As described in Section~\ref{sec:iris_description},
$R_i$ is selected uniformly at random
in the interval that extends from the address of the attacker
to one address before the address of the target $R_i\in [N_i,O_p-1]$,
thus, $x\in [0,o-1]$.
The ${UB}_i$ is $\delta$ addresses away from the address of the attacker.
Assuming $N_i=0$ we will have ${UB}_i=\delta$.
Given a specified value of the random point $x$,
from the attacker's perspective it is equally likely
the target address to be any of the addresses
that succeeds $x$ and precedes $\delta$ (with $\delta$ included).

In Figure~\ref{fig:Cal_P} we illustrate by $O$ the possible
addresses that the target can have given that the randomly
picked address is equal to or equal and less than a specific value $x$.
Given $R_i=x$, in Figure~\ref{fig:Cal_P(O=o|R=x)} and in Figure~\ref{fig:Cal_P(O=o|R<=x)},
we examine what are the possible case(s) for the target to be equal to value $o$
whereas, in Figure~\ref{fig:Cal_P(O<=o|R=x)} and in Figure~\ref{fig:Cal_P(O<=o|R<=x)}
we consider the cases for the target to be equal or less than $o$.
Counting down the number
all the examined events $o$ and
all possible addresses $O$ (that incorporate the number of $o$)
for every $x\in [0,o-1]$, we have that:

\begin{equation}\medmath{
	P(O_p=o|R_i=x)=\frac{1}{\delta-x}}
	\label{eq:prob_P(O=o|R=x)}
\end{equation}

\begin{equation}\medmath{
	P(O_p<=o|R_i=x)=\frac{o-x}{\delta-x}}
	\label{eq:prob_P(O<=o|R=x)}
\end{equation}

\begin{equation}\medmath{
	P(O_p=o|R_i<=x)=\frac{x}{x(\delta-x)+\frac{x(x-1)}{2}}=\frac{2}{2\delta-x-1}}
	\label{eq:prob_P(O=o|R<=x)}
\end{equation}

\begin{equation}\medmath{
	P(O_p<=o|R_i<=x)=\frac{x(o-x)+\frac{x(x-1)}{2}}{x(\delta-x)+\frac{x(x-1)}{2}}
		=\frac{2o-x-1}{2\delta-x-1}}
	\label{eq:prob_P(O<=o|R<=x)}
\end{equation}

From \Cref{eq:prob_P(O=o|R=x),eq:prob_P(O<=o|R=x),eq:prob_P(O=o|R<=x),eq:prob_P(O<=o|R<=x)}
we observe that given the information the attacker has
it is equally likely that the target is any of the possible addresses.
Thus, while considering a powerful adversary
that knows both the $\alpha$ and the $\delta$ parameters
still \alg succeeds in hiding the target from the attacker.

\begin{figure*}[tb]
	\centering
	\subfigure[\alg's convergence for $\delta=1/16$.]{
		\includegraphics[width=0.95\columnwidth]{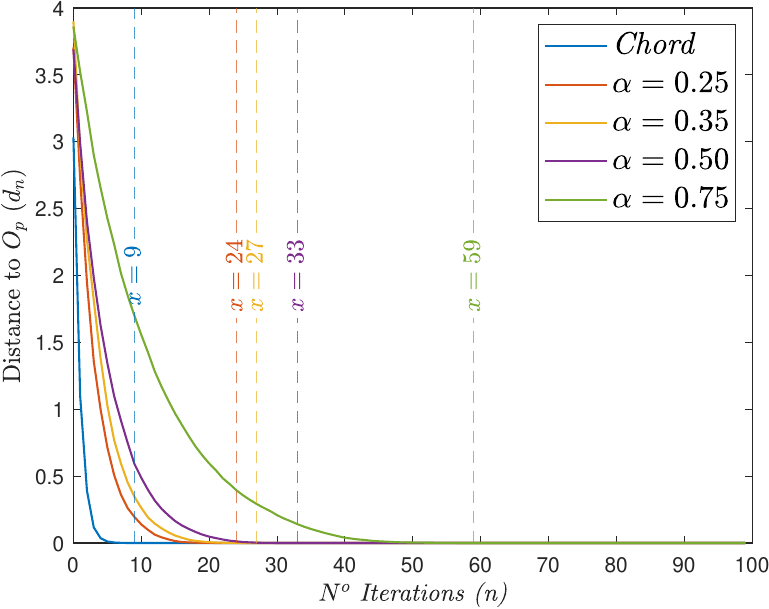}
		\label{fig:convergence_per_alpha}}\hfill
	\subfigure[\alg's convergence for $\alpha=0.35$.]{
		\includegraphics[width=0.95\columnwidth]{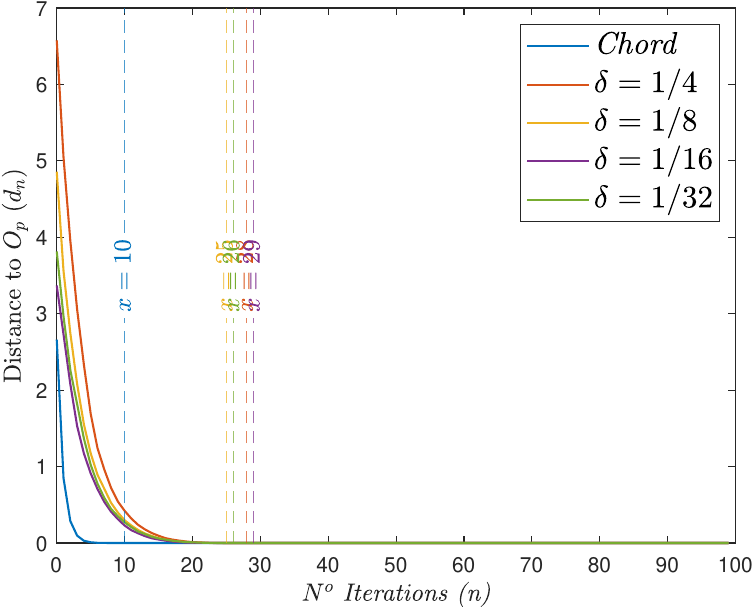}
		\label{fig:convergence_per_delta}}
	\caption{\emph{Comparing \alg's performance for different values of $\alpha$ and $\delta$ parameters.} The x-axis indicates the number of iterations needed to converge to the target, whereas the y-axis indicates the distance the queried nodes have to the target normalised by 1/16 of the address space.}
	\label{fig:convergence}
\end{figure*}

\section{Evaluation}
\label{sec:eval}
We have created a Chord network simulation where we can vary any of
the network parameters independently, and run experiments with any
combination of parameters. We use this to run a large number of
simulations with networks that differ in size, number of nodes,
fraction of adversaries, number of data objects, etc.

We simulate \alg in this environment with a range of choices for the
$\alpha$ and $\delta$ parameters, and analyze how they affect
performance and correctness. We simulate different fractions of
adversaries in the network and analyze the privacy degree we achieve,
accounting for colluding adversaries and perfect ability to guess the
requester's parameters, in accordance with our threat model. We
confirm the probabilistic advantage an attacker has, and find that no
attacker advantage exceeds $\alpha$, thus confirming the \prv of \alg.

We start by describing the setup of our simulation before moving on to
the presentation of the results of our experiments.

\subsection{Simulation Setup}
We simulate \alg using the Matlab programming language.  We model an
address space of $2^{23}$ addresses, on which we position $1000$ nodes
uniformly at random, selecting a fraction $f$ of them to be colluding
adversaries.  We implement Chord \ttt{lookup} and run the network
until a steady state has been reached. Each node keeps a routing table
with $m=23$ other nodes as specified by the Chord protocol. Given such
a network, our implementation selects a requester and a target object
at random from the set of non-attackers, and executes \alg as defined
in Algorithm~\ref{alg:iris}. The requester is free to choose the
$\alpha$ and $\delta$ parameters.

We have open-sourced our simulation code, the evaluation scripts, and
the presented benchmarks as artifacts, and the code is available at
\href{https://github.com/angakt/iris}{https://github.com/angakt/iris}.

Each experiment is run $k$ times, with an entirely new network each
time. This way, our results are independent of the requester and the
target positions in the network, as well as any particular
distribution of attackers. When examining the communication overhead,
\ie the number of iterations \alg needs to terminate, we execute every
experiment $k=100$ times and we report the average distance in every
iteration across all the executions.  When examining the privacy
guarantees \alg provides, we execute every experiment $k=500$ times
and we calculate the minimum ratio of posterior to prior knowledge
across every execution. 

\subsection{Simulation Results}

\begin{figure}[tb]
	\centering
	\includegraphics[width=\columnwidth]{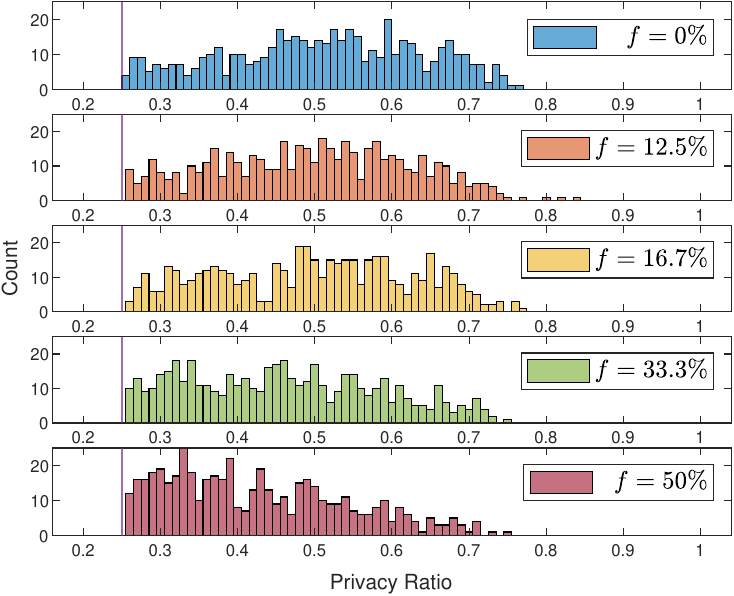}
	\caption{The posterior and prior knowledge rate for different fractions of colluding adversaries.}
	\label{fig:differentFractionsPost2Prior}
\end{figure}

\subsubsection*{Performance}

To understand the performance overheads introduced by \alg we vary the
$\alpha$ and $\delta$ parameters and compare the performance with the
vanilla Chord algorithm. To avoid an impact from differences in
attacker strategy, we run the performance experiments in a network
with no adversaries, \ie $f=0$.

In Figure~\ref{fig:convergence_per_alpha} we report the average
distance to the target on every hop, in a network of $1000$
participants. We evaluate \alg with $\alpha$ equal to $0.25$, $0.35$,
$0.50$, and $0.75$ keeping $\delta$ constant and equal to $2^m/16$.
The performance of \alg is compared to vanilla Chord and averaged over
$k=100$ experiments. The dashed vertical lines indicate the
\emph{maximum} (worst case) number of hops the requester needed to
identify the responsible node for the target object.

We observe that the $\alpha$ parameter has a dominant effect on \alg's
convergence time: the smaller the value of $\alpha$ the faster the
convergence.  This is due to the fact that bigger $\alpha$ values
result in more conservative steps towards the target. This gives away
less information to intermediate nodes, but it also comes with a
performance penalty. This result highlights an important quality of
\alg, namely the fact that the trade-off between performance and
privacy can be tuned to only pay the performance penalty for the
amount of privacy needed.

To asses the performance impact of $\delta$ we keep $\alpha$ constant
($\alpha=0.35$) and vary the $\delta$ parameter. We again run the
experiment 100 times with $f=0$, and the results can be seen in
Figure~\ref{fig:convergence_per_delta}.  We observe only minor
difference in the average number of hops, which matches the prediction
produced by \Cref{eq:hops} where it is clear that the dominant factor
determining \alg's performance is $\alpha$. Changes in the $\delta$
parameter affect the fraction's numerator and have a negligible
influence on the final calculated value. The intuitive explanation for
this is while a larger $\delta$ causes us to start the search further
from the target, we also take larger steps when we are far from the
target, so the overall effect on performance is minor.

\subsubsection*{Query Privacy}
\label{sec:eval_query-privacy}

To validate the query privacy guarantees that \alg achieves with
respect to the privacy notion introduced in \Cref{sec:privacynotion},
we execute \alg by varying the fraction of colluding
adversaries. Following the analysis in
Section~\ref{sec:query-privacy}, we examine the posterior-to-prior
knowledge ratio \alg achieves for each node along the routing
path. Each adversarial node is provided the value of $\alpha$ and
$\delta$ (even though these values would not be available to the
adversaries in practice), and colluding attackers are allowed to
compare values.

For most attackers $\delta$ is an upper bound of where the target
object could be. By knowing $\delta$, the queried nodes that are more
than $\delta$ addresses away will know that, and they do not contribute
to the estimate of later nodes. If they did, the nodes would be wrong
about the target location, so this represents a further advantage for
the attackers that would not exist in practice. By doing this we get
the absolute worst case results for the requester, and therefore a
lower bound on the privacy.

In \Cref{fig:differentFractionsPost2Prior} we illustrate the
\textit{minimum} achieved privacy ratios we get across $k=500$
experiments when we keep $\alpha=0.25$ and $\delta=2^m/4$ and we vary
the fraction of colluding nodes from $f=0\%$ to $f=50\%$.  We notice
that regardless of the fraction of attackers, the privacy ratio does
not drop below $\alpha=0.25$, and is in fact much higher than $\alpha$
in most runs, sometimes as high as 0.7. This confirms what we proved
in \Cref{sec:security}, namely that \alg is an \prvt algorithm.  We
also observe that the greater the fraction of colluding adversaries,
the more frequently we get smaller values of the privacy ratio, i.e.,
the histogram move to the left. However, even for large values of $f$
we remain above $\alpha$ at all times.

\subsubsection*{Attacker Advantage}

To demonstrate the knowledge gain a non-colluding attacker gets from
being used as an intermediate node, we execute \alg with $\alpha=0.75$
and $\delta=2^m/128$, and the fraction of \emph{colluding} adversaries
$f=0$.  We then calculate the distance between an intermediary node
and the target (and to the randomly picked point $R_i$), for every
queried node, and we execute $500$ experiments.

\begin{figure}[tb]
	\centering
	\includegraphics[width=\columnwidth]{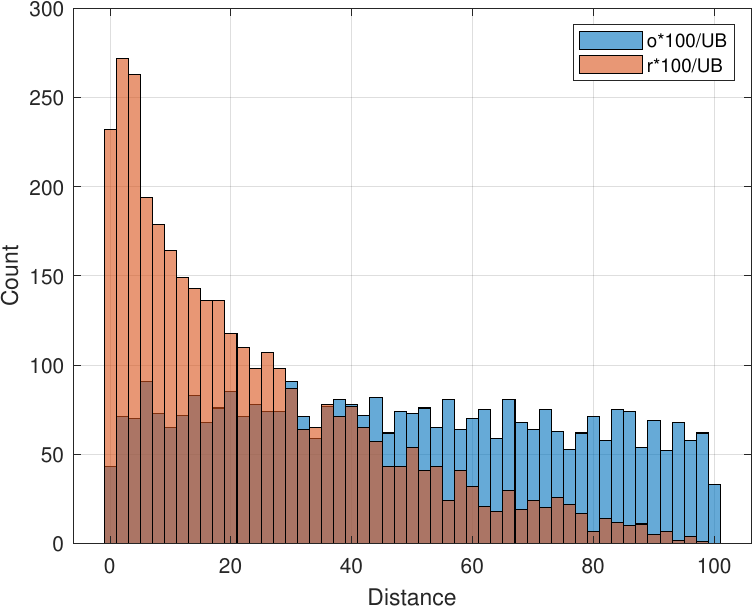}
	\caption{The distance that queried nodes have to the target and to the randomly picked point expressed as a percentage of the $\delta$ value.}
	\label{fig:ProbHistograms}
\end{figure}

In \Cref{fig:ProbHistograms}, we plot a histogram of the normalized
distance to the target (and the random point $R_i$). The distance is
normalized so it corresponds to a percentage of the $\delta$
parameter, since $\delta$ is an upper bound on the distance for almost
all nodes. We observe that the distance to the target follows a
uniform distribution, \ie every node in the interval
$[O_p-\delta,O_p]$ is equally likely to be the target. For the
position of the random point, we observe a right skewed distribution.
This occurs because as the distance between the queried node and the
target gets smaller, the probability of getting a high value for $R_i$
tapers off. Thus, for a uniformly distributed distance to the target
we have more lower than higher values for $R_i$.  A uniform distance
to the target is ideal, since it means that a non-colluding
intermediate node has effectively no information about the location of
the target, other than it is likely to be at most $\delta$ addresses
away.

\begin{figure*}[tb]
	\centering
	\subfigure[$P(O_p=35|R_i=x)$]{
		\resizebox{8.25cm}{!}{\includegraphics[width=\columnwidth]{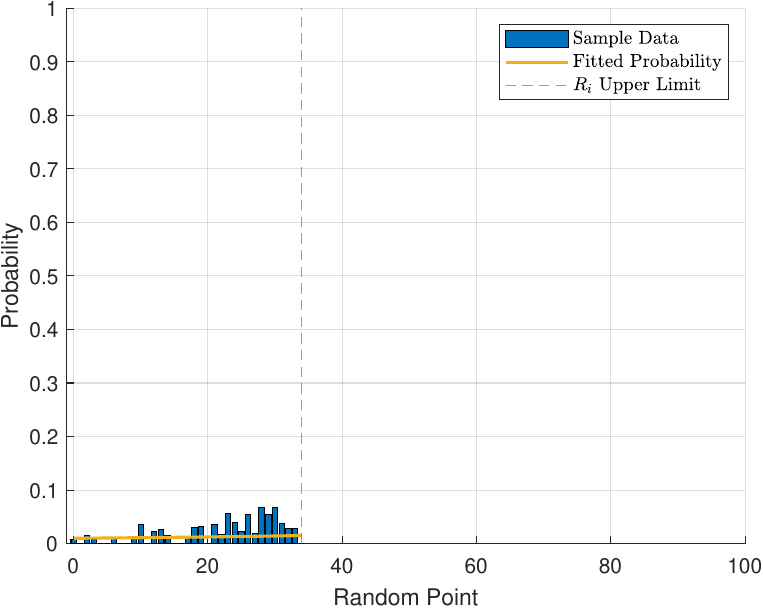}}
		\label{fig:Exec_P(O=o|R=x)}}
	\subfigure[$P(O_p<=35|R_i=x)$]{
		\resizebox{8.25cm}{!}{\includegraphics[width=\columnwidth]{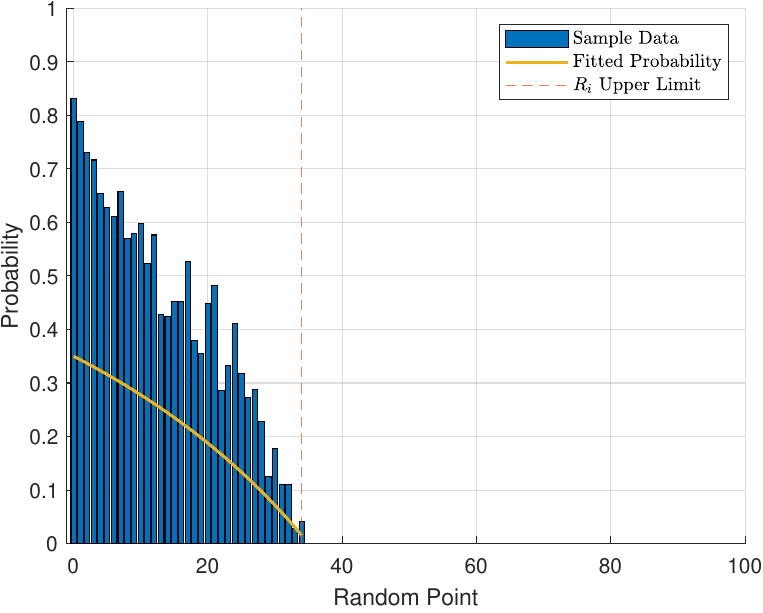}}
		\label{fig:Exec_P(O<=o|R=x)}}
	\subfigure[$P(O_p=35|R_i<=x)$]{
		\resizebox{8.25cm}{!}{\includegraphics[width=\columnwidth]{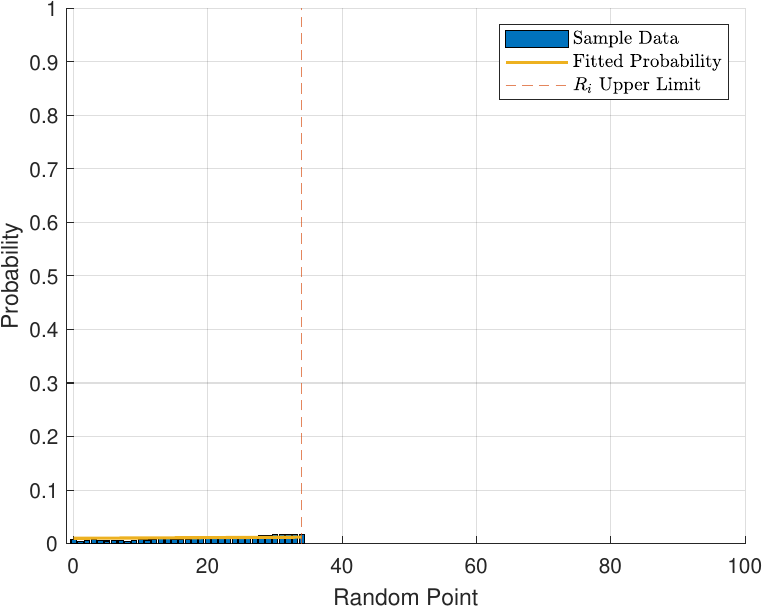}}
		\label{fig:Exec_P(O=o|R<=x)}}
	\subfigure[$P(O_p<=35|R_i<=x)$]{
		\resizebox{8.25cm}{!}{\includegraphics[width=\columnwidth]{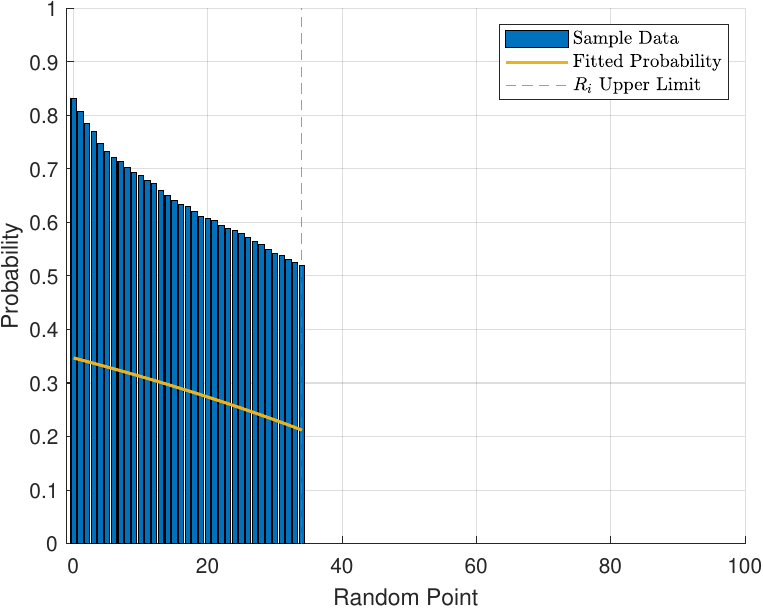}}
		\label{fig:Exec_P(O<=o|R<=x)}}
	\caption{\emph{Probabilities for $O_p=35$:} The probabilities we get from our prototype implementation.}
	\label{fig:Exec_P(O|R)}
\end{figure*}

In \Cref{fig:Exec_P(O|R)}, we validate out implementation of Iris by
plotting the probabilities we get from experiments (in blue) against
the probability expressions we get from Equations
(\ref{eq:prob_P(O=o|R=x)}), (\ref{eq:prob_P(O<=o|R=x)}),
(\ref{eq:prob_P(O=o|R<=x)}) and (\ref{eq:prob_P(O<=o|R<=x)}) (in
yellow). We observe that for
\Cref{fig:Exec_P(O=o|R=x),fig:Exec_P(O=o|R<=x)} the data from our
experiments fully confirm our calculations.  For
\Cref{fig:Exec_P(O<=o|R=x),fig:Exec_P(O<=o|R<=x)} the plots follow the
equation's trend, however, they are higher than expected.  This bias
occurs due to the very common low values we get, that is reflected
when we simultaneously examine more than one value for $O_p$.  In
\Cref{fig:Exec_P(O<=o|R=x)}, we observe that the deviations are
getting smaller as $x$ gets bigger.  In \Cref{fig:Exec_P(O<=o|R<=x)},
we observe that the deviations are maintained as the $x$ gets
bigger. This happens because for this case we examine a range for
values for $x$, thus, any bias in smaller values is inherited to the
bigger ones.  \Cref{fig:Exec_P(O|R)} illustrates that \alg data from
our simulation agrees with the calculated probabilities, or are lower
bounded by them. This acts as a sanity check on the code used for our
simulations and confirms our results described above.

\begin{figure}[tb]
	\centering
	\includegraphics[width=\columnwidth]{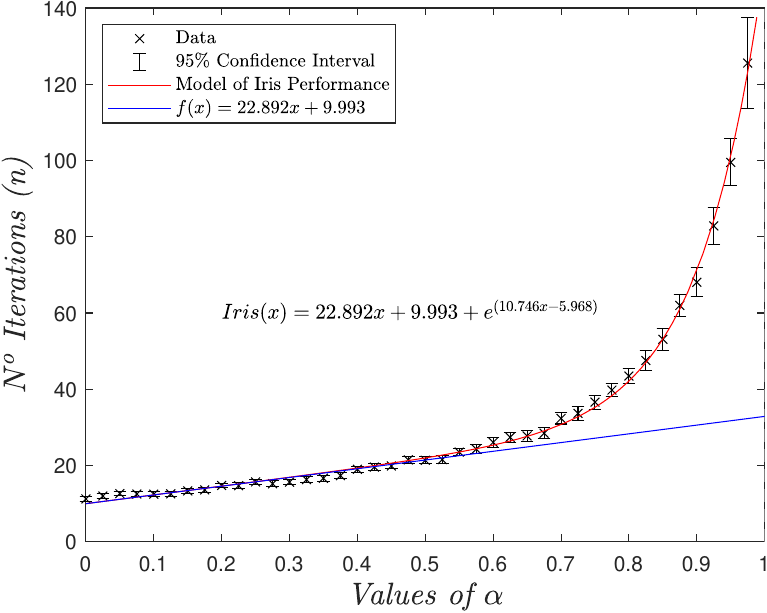}
	\caption{The number of steps that are needed to reach the target for
		different values of the $\alpha$ parameter.}
	\label{fig:alphaValues}
\end{figure}

\section{Related Work}
\label{sec:related}

In this section, we provide background on decentralized privacy schemes
with a focus on the ones applied to Chord.
We start by elaborating on already proposed privacy metrics
before describing other privacy architectures and examining how they compare to \alg.

\subsection{Privacy Metrics}

In the literature, many works have studied the privacy guarantees
of decentralized systems~\cite{systematizing2017}.
To measure the privacy provided in structured P2P architectures
some works extract an anonymity score based
on the size of the anonymity set,
either solely~\cite{tarzan2002,chordleak2004}
or by normalizing it by the best possible value~\cite{anonymity2005borisov}.
In both cases, the calculated score changes between nodes
according to the distance the node has to the target of the request.
Adopting such a metric to measure the query privacy
that is provided as underlined in Section~\ref{sec:privacynotion}
does not reflect the average privacy achieved
but the worst-case scenario.
In~\cite{anonymity2005borisov}, the authors also propose
another anonymity metric based on entropy that takes into consideration the
probabilistic advantage an attacker can have in identifying initiators.
The authors underline that computing the probability of the distribution of events
given an observation can be difficult to apply to complex, dynamic systems.
In~\cite{leakagemodel2007} to calculate the degree of privacy
the authors use the rate between the posterior and the prior entropy
after and before a compromise has occurred.
\prv builds on the last metric, but comparing to that
it necessitates no demanding calculations
by replacing the entropy
with the size of the possible set in which the target can belong.

\subsection{Privacy Architectures}

Structured P2P architectures have been used in
anonymous communication systems such as TOR
as a scalable way to select the
nodes to build anonymous circuits.
This led many works~\cite{secureRouting, ap32004, halo2008, myrmic2006, octapus2012, salsa2006, neblo2006}
to focus on the security of routing, \ie
guarantee an unbiased and correct \ttt{retrieve} process.
\alg tries to enhance the privacy aspect
that is inherently built in Chord so as
privacy guarantees can be achieved
without demanding other infrastructures.

A major line of research focused on
enhancing the anonymity of the sender
and the receiver
in structured P2P architectures~\cite{tarzan2002,MorphMix2002,myrmic2006,nisan2009,torsk2009,shadowwalker2009,xvine2011,WhisperChord2021}.
\alg focuses on a different problem, \ie
hiding the information that is queried from
the intermediate nodes that participate in the routing
while allowing authentication for the participating nodes.

There is a more limited bibliography regarding
the deterrence of user profiling in structured P2P architectures.
The work in~\cite{shardis2012}
split the data and publish every share under a different overlay address,
guaranteeing privacy against an adversary that can capture a small set of shares.
Other works~\cite{queryprivacy2012, chill2021, practical2010} 
organize nodes in quorums; the client uses threshold cryptography to obtain the index of the wanted item without revealing the item and without the individual quorum nodes knowing which item was extracted.
More recently, Peer2PIR~\cite{arxiv2024Peer2PIR} applies private information retrieval (PIR) techniques to limit privacy leakage on peer routing, provider advertisements and content retrieval in the IPFS network, which is based on the Kademlia~\cite{maymounkov2002kademlia} DHT. IPFS is also in the process of performing a privacy upgrade by implementing Double Hashing~\cite{report2023DoubleHashing}. Double Hashing has the requester query for a prefix of the target identifier and receive the records corresponding to any object that matches this prefix, thus guaranteeing $k$-anonymity, where $k$ is the number of objects that match this prefix.
These schemes propose changes in the overlay structure and operations
that have to be followed by all the nodes in the network.
\alg does not demand global changes.

\section{Discussion}
\label{sec:discussion}

In this section we provide further guidance
on the selection of the $\alpha$ and $\delta$ parameters,
and discuss limitations of \alg and its extension to other DHTs.

\subsection{Selection of $\alpha$ and $\delta$ Parameters}

The $\alpha$ and $\delta$ parameters must be selected to be in the
ranges $[0,1)$ and $[0,2^{m}-1]$, respectively. Conceptually, the
bigger the values, the better the provided privacy.  However, as
shown in Figure~\ref{fig:convergence} there is a trade-off between
\alg's privacy and performance, \ie the gain that \alg provides
comes at a cost of increased steps to reach the target.  Thus, it
makes sense to consider what a good selection of the privacy
parameters look like; one that will guarantee a sufficient privacy
level without sacrificing too much performance.

Staring with the $\delta$ parameter, we see in
Figure~\ref{fig:convergence_per_delta} that its value does not
significantly affect the number of iterations until convergence.  This
parameter specifies a minimum distance to the target, and thus
directly controls the anonymity we get against the first queried node
in a group of colluding nodes. This is because such a distance removes
the link between the choice of intermediate node and the target
itself, that vanilla chord would have. Intermediate nodes can no
longer make assumptions about the location of the target, based on the
fact that they are being used as intermediaries. We should pick a
value that contains enough network objects to constitute a good
anonymity set.

As an example consider the Tor network. In Tor we have roughly
$900,000$ onion addresses \cite{TorMetricsOnionServices}.  Assuming a
space of $2^{23}$ addresses that are uniformly distributed, on average
every 9th address will be a valid onion address.  To achieve an
anonymity degree of $k$, we must set $\delta=k*9$.

The selection of the $\alpha$ parameter has a more dominant effect on
performance. This is shown in Figure~\ref{fig:convergence_per_alpha}
but we can see it more clearly in Figure~\ref{fig:alphaValues}, which
illustrates the number of steps \alg needs to converge to the target,
across $100$ experiments altering $\alpha$ with a step of $0.025$ in
the interval $[0,1)$, keeping $\delta=1/16*2^{23}$ and $f=0$.

We observe that the number of steps \alg needs to converge
increases exponentially with respect to $\alpha$, starting at 10,
which is the number of steps that vanilla Chord achieves.  For $\alpha
< 0.7$ the increase in the steps looks almost linear, however above
that, the number of steps starts to increase drastically.  We still
get stronger and stronger privacy guarantees, but with high
performance overheads. For that reason a good practical choice of
$\alpha$ is around $0.7$ which will roughly double the steps (and
search time) of vanilla chord.

Note that each search can be done with a different choice of
parameters, so sensitive searches might need $(.7,500k)$-privacy,
where as normal ones can use no privacy at all.

\subsection{Other P2P Architectures}

\alg can be directly applied to any existing system that implements
the Chord protocol, e.g., \cite{frank2001cfs,nkn2018whitepaper,sit2008usenetdht,
  stribling2006overcite, ajmani02conchord, pirro2016chordgrid,
  labbai2016t2wsn, russ2002dns, kundan2005telephony}. \alg can be used
without any support from the other nodes in the network, even when all
other nodes in the network use the vanilla Chord protocol, which makes
it easy to adopt for privacy conscious clients.

In addition to Chord, several other DHTs exist in the literature, such
as Kademlia~\cite{maymounkov2002kademlia},
Tapestry~\cite{zhao2004tapestry}, Pastry~\cite{rowstron2001pastry} and
CAN~\cite{ratnasamy2001scalable}.  These are, in principle, similar
enough to Chord to provide a way that allows nodes to query their
neighbours and choose the next hop when routing a
message~\cite{gummadi2003impact}.  The general schema that these
protocols offer intuitively suggests that \alg can be applied to these
systems as well.  However, the differences that exist between them,
such as the distance metric they define, may affect the provided
guarantees. For example Chord uses the distance between node addresses
whereas Kademlia uses an XOR metric. We leave a detailed analysis of
the application of \alg to other architectures as future work.

\subsection{Limitations}

As the requester progressively queries nodes that are closer and
closer to the target, intermediate nodes will get a more and more
precise range of possible targets if they are colluding with
previously chosen intermediate nodes. This is an unavoidable
consequence of our very strong threat model, but it is accounted for
in the definition of $(\alpha,\delta)$-privacy and even in the worst
case, the additional information each adversarial node gets is bounded
by $1-\alpha$.

\section{Conclusion}
\label{sec:conclusion}

In this paper, we study the privacy guarantees of a search request
when using Chord. To reason about the query privacy
that a privacy-preserving mechanism provides, we introduce
a new notion called \prv. This privacy notion allows to measure
the privacy level of a request even in the presence of strong colluding adversaries.
We further design \alg, an algorithm that replaces the regular
\ttt{retrieve} algorithm in Chord to allow a requester to conceal the target
of a query from the intermediate nodes that take part in the search.
By performing a thorough security analysis we prove \alg to be both correct and \prvt.
We also perform an empirical analysis using simulations
to evaluate the privacy levels that \alg achieves
and to study the trade-offs our proposal introduces
between the achieved query privacy and performance.
The results confirm our theoretical analysis and indicate a modest communication overhead that can be tuned by the requester based on the privacy level each query demands.

\section*{Acknowledgments}
We want to thank the paper and the artifacts' anonymous reviewers for their time and the valuable feedback they have provided us.



\bibliographystyle{plain}
\bibliography{Iris.bib}

\begin{thebibliography}{10}

\bibitem{ajmani02conchord}
Sameer Ajmani, Dwaine~E. Clarke, Chuang-Hue Moh, and Steven Richman.
\newblock {ConChord}: Cooperative {SDSI} certificate storage and name
  resolution.
\newblock In {\em First International Workshop on Peer-to-Peer Systems
  (IPTPS)}, number 2429 in Lecture Notes in Computer Science, pages 141--154,
  March 2002.

\bibitem{aktypi2022themis}
Angeliki Aktypi, Dimitris Karnikis, Nikos Vasilakis, and Kasper Rasmussen.
\newblock Themis: A secure decentralized framework for microservice interaction
  in serverless computing.
\newblock In {\em Proceedings of the 17th International Conference on
  Availability, Reliability and Security}, ARES '22, New York, NY, USA, 2022.
  ACM.

\bibitem{queryprivacy2012}
Michael Backes, Ian Goldberg, Aniket Kate, and Tomas Toft.
\newblock Adding query privacy to robust dhts.
\newblock In {\em Proceedings of the 7th ACM Symposium on Information, Computer
  and Communications Security}, AsiaCCS '12, page 30–31, New York, NY, USA,
  2012. Association for Computing Machinery.

\bibitem{ingmar2007skademlia}
Ingmar Baumgart and Sebastian Mies.
\newblock S/kademlia: A practicable approach towards secure key-based routing.
\newblock In {\em 2007 International Conference on Parallel and Distributed
  Systems}, pages 1--8, 2007.

\bibitem{benet2014ipfs}
Juan Benet.
\newblock {IPFS} - content addressed, versioned, {P2P} file system.
\newblock {\em CoRR}, abs/1407.3561, 2014.

\bibitem{anonymity2005borisov}
Nikita Borisov and Jason Waddle.
\newblock Anonymity in structured peer-to-peer networks.
\newblock Technical report, Computer Science Division, University of
  California, 2005.

\bibitem{butler2009authentication}
Kevin~R.B. Butler, Sunam Ryu, Patrick Traynor, and Patrick~D. McDaniel.
\newblock Leveraging identity-based cryptography for node id assignment in
  structured p2p systems.
\newblock {\em IEEE Transactions on Parallel and Distributed Systems},
  20(12):1803--1815, 2009.

\bibitem{secureRouting}
Miguel Castro, Peter Druschel, Ayalvadi Ganesh, Antony Rowstron, and Dan~S.
  Wallach.
\newblock Secure routing for structured peer-to-peer overlay networks.
\newblock {\em SIGOPS Oper. Syst. Rev.}, 36(SI):299–314, dec 2003.

\bibitem{neblo2006}
Giuseppe Ciaccio.
\newblock Improving sender anonymity in a structured overlay with imprecise
  routing.
\newblock In George Danezis and Philippe Golle, editors, {\em Privacy Enhancing
  Technologies}, pages 190--207, Berlin, Heidelberg, 2006. Springer Berlin
  Heidelberg.

\bibitem{russ2002dns}
Russ Cox, Athicha Muthitacharoen, and Robert~T. Morris.
\newblock Serving dns using a peer-to-peer lookup service.
\newblock In Peter Druschel, Frans Kaashoek, and Antony Rowstron, editors, {\em
  Peer-to-Peer Systems}, pages 155--165, Berlin, Heidelberg, 2002. Springer
  Berlin Heidelberg.

\bibitem{frank2001cfs}
Frank Dabek, M.~Frans Kaashoek, David Karger, Robert Morris, and Ion Stoica.
\newblock Wide-area cooperative storage with cfs.
\newblock In {\em Proceedings of the Eighteenth ACM Symposium on Operating
  Systems Principles}, SOSP '01, page 202–215, New York, NY, USA, 2001.
  Association for Computing Machinery.

\bibitem{syverson2004tor}
Roger Dingledine, Nick Mathewson, and Paul Syverson.
\newblock Tor: The second-generation onion router.
\newblock In {\em Proceedings of the 13th Conference on USENIX Security
  Symposium - Volume 13}, SSYM'04, page~21, USA, 2004. USENIX Association.

\bibitem{eaton2022onion}
Edward Eaton, Sajin Sasy, and Ian Goldberg.
\newblock Improving the privacy of tor onion services.
\newblock In {\em International Conference on Applied Cryptography and Network
  Security}, pages 273--292. Springer, 2022.

\bibitem{shardis2012}
Benjamin Fabian, Tatiana Ermakova, and Cristian Muller.
\newblock Shardis: A privacy-enhanced discovery service for rfid-based product
  information.
\newblock {\em IEEE Transactions on Industrial Informatics}, 8(3):707--718,
  2012.

\bibitem{tarzan2002}
Michael~J Freedman and Robert Morris.
\newblock Tarzan: A peer-to-peer anonymizing network layer.
\newblock In {\em Proceedings of the 9th ACM Conference on Computer and
  Communications Security}, CCS '02, pages 193--206, New York, NY, USA, 2002.
  ACM.

\bibitem{gummadi2003impact}
K.~Gummadi, R.~Gummadi, S.~Gribble, S.~Ratnasamy, S.~Shenker, and I.~Stoica.
\newblock The impact of dht routing geometry on resilience and proximity.
\newblock In {\em Proceedings of the 2003 Conference on Applications,
  Technologies, Architectures, and Protocols for Computer Communications},
  SIGCOMM '03, page 381–394, New York, NY, USA, 2003. Association for
  Computing Machinery.

\bibitem{halo2008}
Apu Kapadia and Nikos Triandopoulos.
\newblock Halo: High-assurance locate for distributed hash tables.
\newblock In {\em Proceedings of the 16th Network and Distributed System
  Security Symposium}, volume~8 of {\em NDSS '08}, page 142. Citeseer, 2008.

\bibitem{nkn2018whitepaper}
NKN Lab.
\newblock {NKN}: A scalable self-evolving and self-incentivized decentralized
  network.
\newblock Whitepaper, 2018.

\bibitem{storj2018whitepaper}
Storj Lab.
\newblock Storj: A decentralized cloud storage network framework.
\newblock Whitepaper, 2018.

\bibitem{labbai2016t2wsn}
T~Labbai and S~Jothi Prasanna.
\newblock T2wsn: Titivated two-tired chord overlay aiding robustness and
  delivery ratio for wireless sensor networks.
\newblock {\em Journal of Theoretical \& Applied Information Technology},
  91(1), 2016.

\bibitem{maymounkov2002kademlia}
Petar Maymounkov and David Mazieres.
\newblock Kademlia: A peer-to-peer information system based on the xor metric.
\newblock In {\em International Workshop on Peer-to-Peer Systems}, pages
  53--65, Cham, 2002. Springer.

\bibitem{chill2021}
Miti Mazmudar, Stan Gurtler, and Ian Goldberg.
\newblock Do you feel a chill? using pir against chilling effects for
  censorship-resistant publishing.
\newblock In {\em Proceedings of the 20th Workshop on Privacy in the Electronic
  Society}, WPES '21, page 53–57, New York, NY, USA, 2021. Association for
  Computing Machinery.

\bibitem{arxiv2024Peer2PIR}
Miti Mazmudar, Shannon Veitch, and Rasoul~Akhavan Mahdavi.
\newblock Peer2pir: Private queries for ipfs, 2024.

\bibitem{torsk2009}
Jon McLachlan, Andrew Tran, Nicholas Hopper, and Yongdae Kim.
\newblock Scalable onion routing with torsk.
\newblock In {\em Proceedings of the 16th ACM Conference on Computer and
  Communications Security}, CCS '09, page 590–599, New York, NY, USA, 2009.
  Association for Computing Machinery.

\bibitem{report2023DoubleHashing}
Guillaume Michel.
\newblock Ipip-373: Double hash dht spec.
\newblock Technical report, IPFS, 2023.

\bibitem{ap32004}
Alan Mislove, Gaurav Oberoi, Ansley Post, Charles Reis, Peter Druschel, and
  Dan~S. Wallach.
\newblock Ap3: Cooperative, decentralized anonymous communication.
\newblock In {\em Proceedings of the 11th Workshop on ACM SIGOPS European
  Workshop}, EW '04, page 30–es, New York, NY, USA, 2004. Association for
  Computing Machinery.

\bibitem{shadowwalker2009}
Prateek Mittal and Nikita Borisov.
\newblock Shadowwalker: Peer-to-peer anonymous communication using redundant
  structured topologies.
\newblock In {\em Proceedings of the 16th ACM Conference on Computer and
  Communications Security}, CCS '09, page 161–172, New York, NY, USA, 2009.
  ACM.

\bibitem{xvine2011}
Prateek Mittal, Matthew Caesar, and Nikita Borisov.
\newblock X-vine: Secure and pseudonymous routing using social networks.
\newblock In {\em Proceedings of the 2012 Network and Distributed System
  Security Symposium}, NDSS '12, 2012.

\bibitem{salsa2006}
Arjun Nambiar and Matthew Wright.
\newblock Salsa: A structured approach to large-scale anonymity.
\newblock In {\em Proceedings of the 13th ACM Conference on Computer and
  Communications Security}, CCS '06, page 17–26, New York, NY, USA, 2006.
  Association for Computing Machinery.

\bibitem{chordleak2004}
C.W. O'Donnell and V.~Vaikuntanathan.
\newblock Information leak in the chord lookup protocol.
\newblock In {\em Proceedings of the 4th International Conference on
  Peer-to-Peer Computing}, pages 28--35, Aug 2004.

\bibitem{palomar2006authentication}
Esther Palomar, Juan~M. Estevez-Tapiador, Julio~C. Hernandez-Castro, and Arturo
  Ribagorda.
\newblock A p2p content authentication protocol based on byzantine agreement.
\newblock In {\em Emerging Trends in Information and Communication Security},
  pages 60--72, Berlin, Heidelberg, 2006. Springer.

\bibitem{WhisperChord2021}
Andriy Panchenko, Asya Mitseva, and Sara Knabe.
\newblock Whisperchord: Scalable and secure node discovery for overlay
  networks.
\newblock In {\em 2021 IEEE 46th Conference on Local Computer Networks}, LCN
  '21, pages 170--177, 2021.

\bibitem{nisan2009}
Andriy Panchenko, Stefan Richter, and Arne Rache.
\newblock Nisan: Network information service for anonymization networks.
\newblock In {\em Proceedings of the 16th ACM Conference on Computer and
  Communications Security}, CCS '09, page 141–150, New York, NY, USA, 2009.
  Association for Computing Machinery.

\bibitem{pirro2016chordgrid}
Giuseppe Pirrò, Domenico Talia, and Paolo Trunfio.
\newblock A dht-based semantic overlay network for service discovery.
\newblock {\em Future Generation Computer Systems}, 28(4):689--707, 2012.

\bibitem{TorMetricsOnionServices}
Tor Project.
\newblock Tor metrics - onion services.
\newblock \url{https://metrics.torproject.org/hidserv-dir-v3-onions-seen.html},
  2024.
\newblock Accessed: 2024-10-08.

\bibitem{prunster2018holistic}
Bernd Pr{\"u}nster, Dominik Ziegler, Chrisitan Kollmann, and Bojan Suzic.
\newblock A holistic approach towards peer-to-peer security and why proof of
  work won’t do.
\newblock In {\em Security and Privacy in Communication Networks: 14th
  International Conference, SecureComm 2018, Singapore, Singapore, August 8-10,
  2018, Proceedings, Part II}, pages 122--138. Springer, 2018.

\bibitem{ratnasamy2001scalable}
Sylvia Ratnasamy, Paul Francis, Mark Handley, Richard Karp, and Scott Shenker.
\newblock A scalable content-addressable network.
\newblock In {\em Proceedings of the 2001 Conference on Applications,
  Technologies, Architectures, and Protocols for Computer Communications},
  pages 161--172, New York, NY, USA, 2001. ACM.

\bibitem{leakagemodel2007}
Souvik Ray and Zhao Zhang.
\newblock An information-theoretic framework for analyzing leak of privacy in
  distributed hash tables.
\newblock In {\em Proceedings of the 7th IEEE International Conference on
  Peer-to-Peer Computing}, P2P '07, pages 27--36, Sep. 2007.

\bibitem{MorphMix2002}
Marc Rennhard and Bernhard Plattner.
\newblock Introducing morphmix: Peer-to-peer based anonymous internet usage
  with collusion detection.
\newblock In {\em Proceedings of the 2002 ACM Workshop on Privacy in the
  Electronic Society}, WPES '02, page 91–102, New York, NY, USA, 2002.
  Association for Computing Machinery.

\bibitem{rowstron2001pastry}
Antony Rowstron.
\newblock Pastry: Scalable, distributed object location and routing for
  large-scale, persistent peer-to-peer storage utility.
\newblock In {\em IFIP/ACM International Conference on Distributed Plarforms},
  2001.

\bibitem{kundan2005telephony}
Kundan Singh and Henning Schulzrinne.
\newblock Peer-to-peer internet telephony using sip.
\newblock In {\em Proceedings of the International Workshop on Network and
  Operating Systems Support for Digital Audio and Video}, NOSSDAV '05, page
  63–68, New York, NY, USA, 2005. Association for Computing Machinery.

\bibitem{sit2002security}
Emil Sit and Robert Morris.
\newblock Security considerations for peer-to-peer distributed hash tables.
\newblock In {\em Peer-to-Peer Systems}, pages 261--269, Berlin, Heidelberg,
  2002. Springer.

\bibitem{sit2008usenetdht}
Emil Sit, Robert~Tappan Morris, and M~Frans Kaashoek.
\newblock Usenetdht: A low-overhead design for usenet.
\newblock In {\em NSDI}, pages 133--146, 2008.

\bibitem{srivatsa2004vulnerabilities}
Mudhakar Srivatsa and Ling Liu.
\newblock Vulnerabilities and security threats in structured peer-to-peer
  systems: A quantitiative analysis.
\newblock In {\em Proceedings of the 20th Annual Computer Security Applications
  Conference}, volume~10 of {\em ACSAC '04}, pages 252--261, USA, 2004. IEEE
  Computer Society.

\bibitem{stoica2001chord}
Ion Stoica, Robert Morris, David Karger, M.~Frans Kaashoek, and Hari
  Balakrishnan.
\newblock {Chord: A Scalable Peer-to-Peer Lookup Service for Internet
  Applications}.
\newblock {\em ACM SIGCOMM Computer Communication Review}, 31(4):149--160,
  2001.

\bibitem{stribling2006overcite}
Jeremy Stribling, Jinyang Li, Isaac~G Councill, M~Frans Kaashoek, and
  Robert~Tappan Morris.
\newblock Overcite: A distributed, cooperative citeseer.
\newblock In {\em NSDI}, volume~6, pages 11--11, 2006.

\bibitem{systematizing2017}
Carmela Troncoso, Marios Isaakidis, George Danezis, and Harry Halpin.
\newblock Systematizing decentralization and privacy: Lessons from 15 years of
  research and deployments.
\newblock {\em Proceedings on Privacy Enhancing Technologies}, 4:404--426,
  2017.

\bibitem{wallach2003survey}
Dan~S. Wallach.
\newblock A survey of peer-to-peer security issues.
\newblock In {\em Software Security --- Theories and Systems}, pages 42--57,
  Berlin, Heidelberg, 2003. Springer.

\bibitem{myrmic2006}
Peng Wang, Ivan Osipkov, N~Hopper, and Yongdae Kim.
\newblock Myrmic: Secure and robust dht routing.
\newblock Technical report, University of Minnesota, 2006.

\bibitem{octapus2012}
Qiyan Wang and Nikita Borisov.
\newblock Octopus: A secure and anonymous dht lookup.
\newblock In {\em 2012 IEEE 32nd International Conference on Distributed
  Computing Systems}, pages 325--334, June 2012.

\bibitem{practical2010}
Maxwell Young, Aniket Kate, Ian Goldberg, and Martin Karsten.
\newblock Practical robust communication in dhts tolerating a byzantine
  adversary.
\newblock In {\em Proceedings of the 30th IEEE International Conference on
  Distributed Computing Systems}, pages 263--272, June 2010.

\bibitem{zhao2004tapestry}
Ben~Y. Zhao, Ling Huang, Jeremy Stribling, Sean~C. Rhea, Anthony~D. Joseph, and
  John~D. Kubiatowicz.
\newblock Tapestry: A resilient global-scale overlay for service deployment.
\newblock {\em IEEE Journal on Selected Areas in Communications}, 22(1):41--53,
  2004.

\end{thebibliography}

\newpage

\appendices
\section{Artifact Appendix}
\label{app:artifact}

In this work, we introduce \alg, an algorithm that allows nodes that participate in authenticated Chord P2P networks to perform queries without revealing the target of their search to nodes other than the one holding the information. The provided artifacts include the source code of our implementation of the \alg and the Chord algorithms and the code to execute the experiments we have performed to support our claims (together with the datasets we have produced). For completeness, we also include the scripts used to create our plots. The provided artifacts allow peers to reproduce the experiments we present in the paper and build upon them, inspiring further research and development in this area.

\subsection{Artifact Structure}

In the repository we provide the code and the data we use in the paper.
As a fist step, users can run the code to execute the \alg algorithm, and to collect execution data.
Subsequently, the users can run the provided plot scripts to recreate the figures used in the paper.
The network generation is randomized, so to fully reproduce the results we present in the paper, we share the data we generated and used for our plotting.

\begin{figure}[tb]
	\centering
	\includegraphics[width=1\columnwidth]{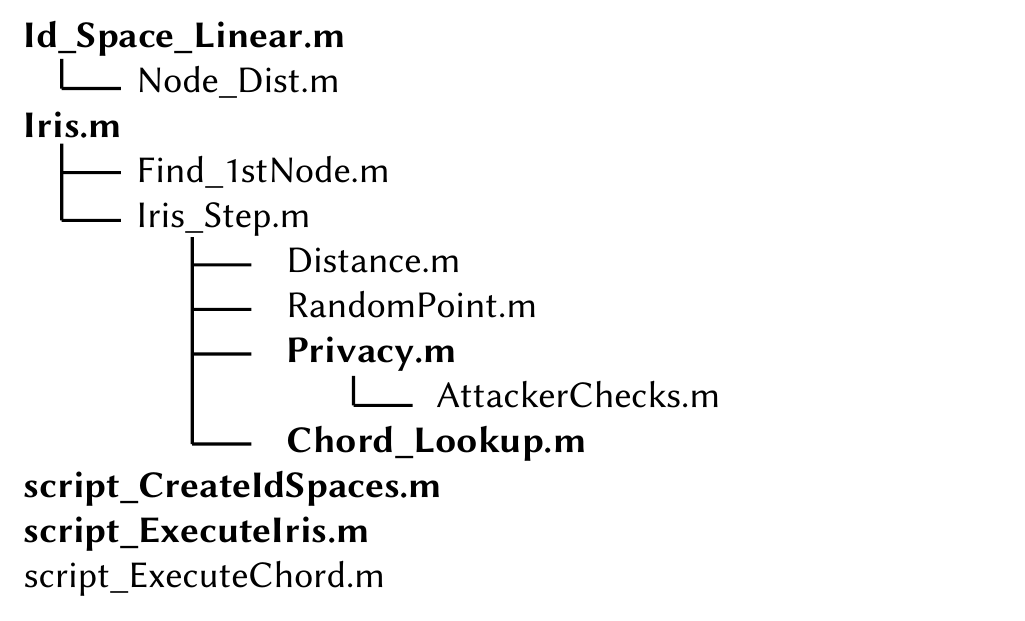}
	\caption{\emph{The schema of the Iris code base}.}
	\label{fig:ArtifactSchema}
\end{figure}

In Figure~\ref{fig:ArtifactSchema} we provide the schema of the coding files in the repository.
The majority of the work is done in these four files:

\begin{itemize}
\item \ttt{Id\_Space\_Linear.m} creates a circular id space
with a number of participating nodes placed uniformly at random.
A fraction of them, \ie $0, \frac{1}{2}, \frac{1}{3}, \frac{1}{8}, \ldots$, are chosen as colluding adversaries.
%
%
\item \ttt{Iris.m} implements the \alg algorithm as described in Section~\ref{sec:iris} of the paper.
This function performs an iterative search for a target object $O_p$, 
initiated by a requester $N_r$ keeping the privacy parameters $\alpha$ and $\delta$
constant during the search.

\item \ttt{Privacy.m} calculates the loss of privacy at every node that is queried in a search. This loss
is proportional to the ratio of the \textit{posterior} and the \textit{prior} knowledge about the target object,
as specified in Section~\ref{sec:privacynotion} of our paper.
  
\item \ttt{Chord\_Lookup.m} implements the Chord \ttt{lookup} protocol,
as described in Section~\ref{sec:bg} of our paper. Given a specified target object, the code
returns the address of the next node to be queried.
\end{itemize}

\subsection{Set Up}

\subsubsection{Access}
The complete artifacts and most recent version of the code can be accessed at \href{https://github.com/angakt/iris}{https://github.com/angakt/iris}.
A snapshot of the release (based on which the artifacts evaluation has been performed) has also been uploaded at Zenodo and can be accessed at \href{https://doi.org/10.5281/zenodo.14251874}{https://doi.org/10.5281/zenodo.14251874}.

\subsubsection{Hardware Dependencies}
Any computer that can run Matlab or GNU Octave.
For our implementation we use a computer with
an Intel Core i7-4820K processor and 16GB installed RAM memory.

\subsubsection{Software Dependencies}
The code can be executed on any operation system that supports Matlab or Octave, e.g., Windows, MacOS, or Ubuntu.
We test our code on Windows 10,  Education edition, version 22H2,
running the Matlab version R2023a with an academic licence.
However, as the code does not use any special libraries, GNU Octave,
an open-source alternative to Matlab, is also sufficient to run the scripts.
For convenience we link to a GNU Octave docker image
available at \href{https://github.com/gnu-octave/docker}{https://github.com/gnu-octave/docker}

\subsection{Claims}

The provided code is used to evaluate the proposed algorithm, \alg.
More specifically the experiments performed help us attest:

\begin{itemize}
	\item [C1] \alg's correctness: the algorithm converges to the target address.
	\item [C2] \alg's privacy guarantees: the algorithm is \prvt, \ie the privacy ratio against any queried node and colluding adversary does not drop below $\alpha$, and the algorithm does not provide an advantage to the attacker's guess.

\end{itemize}

\subsection{Execution}

In the paper we perform five different experiments. The first two experiments
presented in Figure~\ref{fig:convergence} of the paper, focus on the evaluation of the performance cost
that is introduced by \alg, \ie the extra hops that a query needs to perform.
The third experiment presented in Figure~\ref{fig:differentFractionsPost2Prior}, attests that \alg
is an \prvt algorithm, whereas the last two experiments
illustrated in Figure~\ref{fig:ProbHistograms} and in Figure~\ref{fig:Exec_P(O|R)} examines the
attacker's advantage when executing \alg.

\subsubsection{Network Generation}
The preliminary step for all the experiments is the creation
of a number of different networks.
These networks are generated with the \ttt{script\_CreateIdSpaces.m} script.

The script creates 500 different networks of 1000 nodes each
with $2^{23}$ number of addresses.
These three parameters are hardcoded in the script but
can be changed based on user's needs
by changing lines 7, 11 and 12, respectively.
After executing this command 500 mat files,
each one containing one initialised network, are created
and saved under the folder
\ttt{./experiments/networks/1000\_nodes/}.
After completing this preliminary step
we can proceed with the execution of the experiments.

\subsubsection{Experiments}
For all the experiments we make use of the \ttt{script\_ExecuteIris.m}
script. This script allows us to run \alg a specified number of times,
specifying the $\alpha$ and $\delta$ parameters, and the fraction of
colluding adversaries, across every set of experiments.

For every execution, the script uses a different network of 1000 nodes
by loading a new network file
from the \ttt{./experiments/networks/1000\_nodes/} folder.
For every execution the script selects the address of the target and the requester uniformly at random,
checking that the requester is not among the colluding nodes.

The experiment parameters are embedded in the script
thus for every new experiment a few lines needs to be changed to generate the required data for a particular experiment.

To reproduce our experiments and results, the following lines need to
be changed:

\begin{itemize}
	\item line 12 specifies the number of performed experiments of every set.
	\item lines 28, 29 and 34 specify the number of participating nodes in the network,
	the number of the id space addresses and the folder under which the mat file
	with the initialised network is saved, respectively.
	\item line 42 defines the $\alpha$ parameter.
	\item line 47 defines the $\delta$ parameter.
	\item line 53 defines the fraction of colluding nodes.
	(Recall that the colluding nodes are specified in the attackers variable in the mat file of the address space.)
	\item line 131 defines the name of the file to be saved with the experiments data.
\end{itemize}

To avoid an error-prone reproducibility of the performed experiments,
we provide the parametrized \ttt{script\_ExecuteIris.m} scripts that are to be used for every experiment.
In the next section, we report their use together with further details regarding the execution of the experiments.

\subsubsection{Graphs}

Finally, to reproduce the graphs used in the paper, the data from the
experiments can be plotted with the scripts found in the \ttt{./experiments/results/}
folder. These are standard plots in either Matlab or R and we do not
consider these part of our contribution, but we include them for
completeness.

\balance

\subsection{Evaluation}

\textit{[Preparation]}

All the experiments in the paper were executed using networks with 1000 nodes placed on an address space with $2^{23}$ addresses.

\subsubsection{Experiment (E1)}
[Figure~\ref{fig:convergence_per_alpha}] [1 human-minute + 1 compute-minute]:
In this experiment we examine \alg's performance for different values of the $\alpha$ parameter,
to support our first claim.

\textit{[Execution]}

1) Run the script \path{results\fig_DistancesPerAlpha\script_ExecuteIris.m}, this will execute \alg setting
$\alpha$ equal to 0.25, 0.35, 0.5 and 0.75, producing 4 csv and 4 mat files.
For each value of the $\alpha$ parameter we execute 100 experiments.
Apart from $\alpha$ the rest parameters remain stable, $\delta=1/16*address\_space$ and $f=0$.

2) Run the script \path{results\fig_DistancesPerAlpha\script_ExecuteChord.m}, this will produce 1 csv file (data\_a1.csv) that
contains the results when executing Chord using for comparison reasons the targets and the requesters of one of the other mat files.

\textit{[Graph]}

Run the \path{results\fig_DistancesPerAlpha\script_PlotDistancesPerAlpha.m} to plot the average distances to the target
for each $\alpha$ value. The plot needs to be executed in the same folder with the data produced above.

\subsubsection{Experiment (E2)}
[Figure~\ref{fig:convergence_per_delta}] [1 human-minute + 1 compute-minute]:
This experiment is also related to our first claim examining
\alg convergence for different values of the $\delta$ parameter.

\textit{[Execution]}

1) Run the script \path{results\fig_DistancesPerDelta\script_ExecuteIris.m}, this will execute \alg setting
$\delta$ equal to 1/4, 1/8, 1/16 and 1/32 of the address space, producing 4 csv and 4 mat files.
For each value of the $\delta$ parameter we execute 100 experiments.
Apart from $\delta$ the rest parameters remain stable, $\alpha=0.35$ and $f=0$.

2) Run the script \path{results\fig_DistancesPerDelta\script_ExecuteChord.m}, this will produce 1 csv file (data\_a1.csv) that
contains the results when executing Chord using for comparison reasons the targets and the requesters of one of the other mat files.

\textit{[Graph]}

Run the \path{results\fig_DistancesPerDelta\script_PlotDistancesPerDelta.m} to plot the average distances to the target
for each $\delta$ value. The plot needs to be executed in the same folder with the data produced above.

\textit{[Preparation]}

For the experiments in Figures~\ref{fig:differentFractionsPost2Prior},~\ref{fig:ProbHistograms} and~\ref{fig:Exec_P(O|R)}
we alter \alg so as to focus solely on the nodes that have a correct
estimation regarding the target. Thus, we need to comment lines 27-31 and uncomment lines 35-40 in the \path{Iris.m} file.
The next three experiments support our second major claim.

\subsubsection{Experiment (E3)}
[Figure~\ref{fig:differentFractionsPost2Prior}] [1 human-minute + 5 compute-minutes]:

\textit{[Execution]}

1) Run the script \path{results\fig_PrivacyPerAttackers\script_ExecuteIris.m}, this will execute \alg setting
the $f$ value equal to 0, 1/2, 1/3, 1/6 and 1/8, producing 5 mat files.
For each $f$ value we execute 500 experiments.
Apart from $f$ the rest parameters remain stable, $\alpha=0.25$ and $\delta=1/4*address\_space$.

2) Run the script \path{results\fig_PrivacyPerAttackers\script_FindMinPrivacyRatio.m}, the script loads the privacy data of the 500 experiments of
each $f$ value and finds the min privacy ratio of every experiment saving the data to 5 csv files.

\textit{[Graph]}

Run the script \path{results\fig_PrivacyPerAttackers\script_PlotMinPrivacyRatioPerAttackers} to plot the minimum acquired privacy ratios as histograms.

\subsubsection{Experiment (E4)}
[Figure~\ref{fig:ProbHistograms}] [1 human-minute + 1 compute-minute]:

\textit{[Execution]}

1) Run the script \path{results\fig_Probabilities\fig_DistancesNormalizedByDelta\script_ExecuteIris.m}, this will execute \alg 500 times 
with $\alpha=0.75$, $\delta=1/128*address\_space$ and $f=0$, producing 1 mat file.

\textit{[Graph]}

Run the \path{results\fig_Probabilities\fig_DistancesNormalizedByDelta\script_PlotDistancesNormalizedByDelta.m} to plot the histogram of the results.

\subsubsection{Experiment (E5)}
[Figure~\ref{fig:Exec_P(O|R)}] [1 human-minute + 1 compute-minute]:

\textit{[Execution]}

1) The script \path{results\fig_Probabilities\fig_DistancesNormalizedByDelta\script_PlotDistancesNormalizedByDelta.m} from the previous step,
produces 2 csv files with the distances the queried node has to the target and to the randomly picked address.
If we do not want the plotting but only to extract the two csv files that are necessary for the fifth experiment,
we have to comment lines 28-45.

\textit{[Graphs]}

To plot the probabilities we execute the scripts in the \path{results\fig_Probabilities\fig_ConditionalProbabilities} folder.
Each script corresponds to one subfigure. We can alter the examined $x$ value by changing line 9.

\end{document}